\documentclass[reprint,aps,prd,twocolumn,superscriptaddress,showpacs,preprintnumbers,amsmath,amssymb,floatfix,merge]{revtex4-1}
\usepackage{graphicx}  
\usepackage[mathlines]{lineno}
\usepackage{dcolumn}   
\usepackage{bm}        
\usepackage{amssymb}   
\usepackage{xcolor}
\usepackage{verbatim}
\usepackage{tabularx} 
\usepackage[colorlinks,linkcolor=blue,citecolor=blue,urlcolor=blue]{hyper ref}
\usepackage{array}
\usepackage{nicefrac}

\usepackage{pdfpages}

\newcolumntype{b}{X}
\newcolumntype{m}{>{\hsize=.8\hsize}X}
\newcolumntype{s}{>{\hsize=.7\hsize}X}

\def\gev{GeV/\textit{c}$^2$}

\begin{document}



\title{Dark matter effective field theory scattering in direct detection experiments}
\affiliation{Division of Physics, Mathematics, \& Astronomy, California Institute of Technology, Pasadena, CA 91125, USA}
\affiliation{Institute for Particle Physics Phenomenology, Department of Physics, Durham University, Durham, UK}
\affiliation{Fermi National Accelerator Laboratory, Batavia, IL 60510, USA}
\affiliation{Lawrence Berkeley National Laboratory, Berkeley, CA 94720, USA}
\affiliation{Department of Physics, Massachusetts Institute of Technology, Cambridge, MA 02139, USA}
\affiliation{Pacific Northwest National Laboratory, Richland, WA 99352, USA}
\affiliation{Department of Physics, Queen's University, Kingston ON, Canada K7L 3N6}
\affiliation{Department of Physics, Santa Clara University, Santa Clara, CA 95053, USA}
\affiliation{SLAC National Accelerator Laboratory/Kavli Institute for Particle Astrophysics and Cosmology, Menlo Park, CA 94025, USA}
 \affiliation{Department of Physics, South Dakota School of Mines and Technology, Rapid City, SD 57701, USA}	
\affiliation{Department of Physics, Southern Methodist University, Dallas, TX 75275, USA}
\affiliation{Department of Physics, Stanford University, Stanford, CA 94305, USA}
\affiliation{Department of Physics, Syracuse University, Syracuse, NY 13244, USA}
\affiliation{Department of Physics and Astronomy, and the Mitchell Institute for Fundamental Physics and Astronomy, Texas A\&M University, College Station, TX 77843, USA}
\affiliation{Departamento de F\'{\i}sica Te\'orica and Instituto de F\'{\i}sica Te\'orica UAM/CSIC, Universidad Aut\'onoma de Madrid, 28049 Madrid, Spain}
\affiliation{Department of Physics \& Astronomy, University of British Columbia, Vancouver,  BC  V6T 1Z1, Canada}
\affiliation{Department of Physics, University of California, Berkeley, CA 94720, USA}
\affiliation{Department of Physics, University of California, Santa Barbara, CA 93106, USA}
\affiliation{Department of Physics, University of Colorado Denver, Denver, CO 80217, USA}
\affiliation{Department of Physics, University of Evansville, Evansville, IN 47722, USA}
\affiliation{Department of Physics, University of Florida, Gainesville, FL 32611, USA}
\affiliation{Department of Physics, University of Illinois at Urbana-Champaign, Urbana, IL 61801, USA}
\affiliation{School of Physics \& Astronomy, University of Minnesota, Minneapolis, MN 55455, USA}
\affiliation{Department of Physics, University of South Dakota, Vermillion, SD 57069, USA}

\author{K.~Schneck} \email{Corresponding author: kschneck@stanford.edu} \affiliation{SLAC National Accelerator Laboratory/Kavli Institute for Particle Astrophysics and Cosmology, Menlo Park, CA 94025, USA}
\author{B.~Cabrera} \affiliation{Department of Physics, Stanford University, Stanford, CA 94305, USA}
\author{D.G.~Cerde\~no} \affiliation{Institute for Particle Physics Phenomenology, Department of Physics,
Durham University, Durham, UK}
\author{V.~Mandic} \affiliation{School of Physics \& Astronomy, University of Minnesota, Minneapolis, MN 55455, USA}
\author{H.~E.~Rogers} \affiliation{School of Physics \& Astronomy, University of Minnesota, Minneapolis, MN 55455, USA}
\author{R.~Agnese} \affiliation{Department of Physics, University of Florida, Gainesville, FL 32611, USA}	
\author{A.J.~Anderson} \affiliation{Department of Physics, Massachusetts Institute of Technology, Cambridge, MA 02139, USA}
\author{M.~Asai} \affiliation{SLAC National Accelerator Laboratory/Kavli Institute for Particle Astrophysics and Cosmology, Menlo Park, CA 94025, USA}
\author{D.~Balakishiyeva} \affiliation{Department of Physics, University of Florida, Gainesville, FL 32611, USA}	
\author{D.~Barker} \affiliation{School of Physics \& Astronomy, University of Minnesota, Minneapolis, MN 55455, USA}	
\author{R.~Basu~Thakur~} \affiliation{Fermi National Accelerator Laboratory, Batavia, IL 60510, USA}\affiliation{Department of Physics, University of Illinois at Urbana-Champaign, Urbana, IL 61801, USA}	
\author{D.A.~Bauer} \affiliation{Fermi National Accelerator Laboratory, Batavia, IL 60510, USA}	
\author{J.~Billard} \affiliation{Department of Physics, Massachusetts Institute of Technology, Cambridge, MA 02139, USA}	
\author{A.~Borgland}\affiliation{SLAC National Accelerator Laboratory/Kavli Institute for Particle Astrophysics and Cosmology, Menlo Park, CA 94025, USA}
\author{D.~Brandt}\affiliation{SLAC National Accelerator Laboratory/Kavli Institute for Particle Astrophysics and Cosmology, Menlo Park, CA 94025, USA}
\author{P.L.~Brink} \affiliation{SLAC National Accelerator Laboratory/Kavli Institute for Particle Astrophysics and Cosmology, Menlo Park, CA 94025, USA}
\author{R.~Bunker} \affiliation{Department of Physics, South Dakota School of Mines and Technology, Rapid City, SD 57701, USA}	
\author{D.O.~Caldwell} \affiliation{Department of Physics, University of California, Santa Barbara, CA 93106, USA}	
\author{R.~Calkins} \affiliation{Department of Physics, Southern Methodist University, Dallas, TX 75275, USA}	
\author{H.~Chagani} \affiliation{School of Physics \& Astronomy, University of Minnesota, Minneapolis, MN 55455, USA}	
\author{Y.~Chen} \affiliation{Department of Physics, Syracuse University, Syracuse, NY 13244, USA}	
\author{J.~Cooley} \affiliation{Department of Physics, Southern Methodist University, Dallas, TX 75275, USA}	
\author{B.~Cornell} \affiliation{Division of Physics, Mathematics, \& Astronomy, California Institute of Technology, Pasadena, CA 91125, USA}	
\author{C.H.~Crewdson} \affiliation{Department of Physics, Queen's University, Kingston ON, Canada K7L 3N6}	
\author{P.~Cushman} \affiliation{School of Physics \& Astronomy, University of Minnesota, Minneapolis, MN 55455, USA}	
\author{M.~Daal} \affiliation{Department of Physics, University of California, Berkeley, CA 94720, USA}	
\author{P.C.F.~Di~Stefano} \affiliation{Department of Physics, Queen's University, Kingston ON, Canada K7L 3N6}	
\author{T.~Doughty} \affiliation{Department of Physics, University of California, Berkeley, CA 94720, USA}
\author{L.~Esteban} \affiliation{Departamento de F\'{\i}sica Te\'orica and Instituto de F\'{\i}sica Te\'orica UAM/CSIC, Universidad Aut\'onoma de Madrid, 28049 Madrid, Spain} 	
\author{S.~Fallows} \affiliation{School of Physics \& Astronomy, University of Minnesota, Minneapolis, MN 55455, USA}	
\author{E.~Figueroa-Feliciano} \affiliation{Department of Physics, Massachusetts Institute of Technology, Cambridge, MA 02139, USA}	
\author{G.L.~Godfrey} \affiliation{SLAC National Accelerator Laboratory/Kavli Institute for Particle Astrophysics and Cosmology, Menlo Park, CA 94025, USA}
\author{S.R.~Golwala} \affiliation{Division of Physics, Mathematics, \& Astronomy, California Institute of Technology, Pasadena, CA 91125, USA}	
\author{J.~Hall} \affiliation{Pacific Northwest National Laboratory, Richland, WA 99352, USA}	
\author{H.R.~Harris} \affiliation{Department of Physics and Astronomy, and the Mitchell Institute for Fundamental Physics and Astronomy, Texas A\&M University, College Station, TX 77843, USA}
\author{T.~Hofer} \affiliation{School of Physics \& Astronomy, University of Minnesota, Minneapolis, MN 55455, USA}	
\author{D.~Holmgren} \affiliation{Fermi National Accelerator Laboratory, Batavia, IL 60510, USA}	
\author{L.~Hsu} \affiliation{Fermi National Accelerator Laboratory, Batavia, IL 60510, USA}	
\author{M.E.~Huber} \affiliation{Department of Physics, University of Colorado Denver, Denver, CO 80217, USA}	
\author{D.M.~Jardin} \affiliation{Department of Physics, Southern Methodist University, Dallas, TX 75275, USA}	
\author{A.~Jastram} \affiliation{Department of Physics and Astronomy, and the Mitchell Institute for Fundamental Physics and Astronomy, Texas A\&M University, College Station, TX 77843, USA}
\author{O.~Kamaev} \affiliation{Department of Physics, Queen's University, Kingston ON, Canada K7L 3N6}	
\author{B.~Kara} \affiliation{Department of Physics, Southern Methodist University, Dallas, TX 75275, USA}	
\author{M.H.~Kelsey} \affiliation{SLAC National Accelerator Laboratory/Kavli Institute for Particle Astrophysics and Cosmology, Menlo Park, CA 94025, USA}
\author{A.~Kennedy} \affiliation{School of Physics \& Astronomy, University of Minnesota, Minneapolis, MN 55455, USA}		
\author{A.~Leder} \affiliation{Department of Physics, Massachusetts Institute of Technology, Cambridge, MA 02139, USA}	
\author{B.~Loer} \affiliation{Fermi National Accelerator Laboratory, Batavia, IL 60510, USA}	
\author{E.~Lopez~Asamar} \affiliation{Departamento de F\'{\i}sica Te\'orica and Instituto de F\'{\i}sica Te\'orica UAM/CSIC, Universidad Aut\'onoma de Madrid, 28049 Madrid, Spain} 	
\author{P.~Lukens} \affiliation{Fermi National Accelerator Laboratory, Batavia, IL 60510, USA}	
\author{R.~Mahapatra} \affiliation{Department of Physics and Astronomy, and the Mitchell Institute for Fundamental Physics and Astronomy, Texas A\&M University, College Station, TX 77843, USA}	
\author{K.A.~McCarthy} \affiliation{Department of Physics, Massachusetts Institute of Technology, Cambridge, MA 02139, USA}	
\author{N.~Mirabolfathi} \affiliation{Department of Physics and Astronomy, and the Mitchell Institute for Fundamental Physics and Astronomy, Texas A\&M University, College Station, TX 77843, USA}	
\author{R.A.~Moffatt} \affiliation{Department of Physics, Stanford University, Stanford, CA 94305, USA}	
\author{J.D.~Morales~Mendoza} \affiliation{Department of Physics and Astronomy, and the Mitchell Institute for Fundamental Physics and Astronomy, Texas A\&M University, College Station, TX 77843, USA}	
\author{S.M.~Oser} \affiliation{Department of Physics \& Astronomy, University of British Columbia, Vancouver,  BC  V6T 1Z1, Canada}	
\author{K.~Page} \affiliation{Department of Physics, Queen's University, Kingston ON, Canada K7L 3N6}	
\author{W.A.~Page} \affiliation{Department of Physics \& Astronomy, University of British Columbia, Vancouver,  BC  V6T 1Z1, Canada}	
\author{R.~Partridge} \affiliation{SLAC National Accelerator Laboratory/Kavli Institute for Particle Astrophysics and Cosmology, Menlo Park, CA 94025, USA}
\author{M.~Pepin} \affiliation{School of Physics \& Astronomy, University of Minnesota, Minneapolis, MN 55455, USA}	
\author{A.~Phipps} \affiliation{Department of Physics, University of California, Berkeley, CA 94720, USA}	
\author{K.~Prasad} \affiliation{Department of Physics and Astronomy, and the Mitchell Institute for Fundamental Physics and Astronomy, Texas A\&M University, College Station, TX 77843, USA}
\author{M.~Pyle} \affiliation{Department of Physics, University of California, Berkeley, CA 94720, USA}	
\author{H.~Qiu} \affiliation{Department of Physics, Southern Methodist University, Dallas, TX 75275, USA}	
\author{W.~Rau} \affiliation{Department of Physics, Queen's University, Kingston ON, Canada K7L 3N6}	
\author{P.~Redl} \affiliation{Department of Physics, Stanford University, Stanford, CA 94305, USA}	
\author{A.~Reisetter} \affiliation{Department of Physics, University of Evansville, Evansville, IN 47722, USA}	
\author{Y.~Ricci} \affiliation{Department of Physics, Queen's University, Kingston ON, Canada K7L 3N6}	
\author{A.~Roberts} \affiliation{Department of Physics, University of South Dakota, Vermillion, SD 57069, USA}
\author{T.~Saab} \affiliation{Department of Physics, University of Florida, Gainesville, FL 32611, USA}	
\author{B.~Sadoulet} \affiliation{Department of Physics, University of California, Berkeley, CA 94720, USA}\affiliation{Lawrence Berkeley National Laboratory, Berkeley, CA 94720, USA}	
\author{J.~Sander} \affiliation{Department of Physics, University of South Dakota, Vermillion, SD 57069, USA}
\author{R.W.~Schnee} \affiliation{Department of Physics, South Dakota School of Mines and Technology, Rapid City, SD 57701, USA}	
\author{S.~Scorza} \affiliation{Department of Physics, Southern Methodist University, Dallas, TX 75275, USA}	
\author{B.~Serfass} \affiliation{Department of Physics, University of California, Berkeley, CA 94720, USA}	 
\author{B.~Shank} \affiliation{Department of Physics, Stanford University, Stanford, CA 94305, USA}	
\author{D.~Speller} \affiliation{Department of Physics, University of California, Berkeley, CA 94720, USA}	
\author{D.~Toback} \affiliation{Department of Physics and Astronomy, and the Mitchell Institute for Fundamental Physics and Astronomy, Texas A\&M University, College Station, TX 77843, USA}
\author{S.~Upadhyayula}\affiliation{Department of Physics and Astronomy, and the Mitchell Institute for Fundamental Physics and Astronomy, Texas A\&M University, College Station, TX 77843, USA}
\author{A.N.~Villano} \affiliation{School of Physics \& Astronomy, University of Minnesota, Minneapolis, MN 55455, USA}	
\author{B.~Welliver} \affiliation{Department of Physics, University of Florida, Gainesville, FL 32611, USA}	
\author{J.S.~Wilson} \affiliation{Department of Physics and Astronomy, and the Mitchell Institute for Fundamental Physics and Astronomy, Texas A\&M University, College Station, TX 77843, USA}	
\author{D.H.~Wright} \affiliation{SLAC National Accelerator Laboratory/Kavli Institute for Particle Astrophysics and Cosmology, Menlo Park, CA 94025, USA}	
\author{X.~Yang} \affiliation{Department of Physics, University of South Dakota, Vermillion, SD 57069, USA}	
\author{S.~Yellin} \affiliation{Department of Physics, Stanford University, Stanford, CA 94305, USA}	
\author{J.J.~Yen} \affiliation{Department of Physics, Stanford University, Stanford, CA 94305, USA}	
\author{B.A.~Young} \affiliation{Department of Physics, Santa Clara University, Santa Clara, CA 95053, USA}	
\author{J.~Zhang} \affiliation{School of Physics \& Astronomy, University of Minnesota, Minneapolis, MN 55455, USA}

\collaboration{SuperCDMS Collaboration} 
\noaffiliation

\begin{abstract}

We examine the consequences of the effective field theory (EFT) of dark matter-nucleon scattering for current and proposed direct detection experiments.  Exclusion limits on EFT coupling constants computed using the optimum interval method are presented for SuperCDMS Soudan, CDMS II, and LUX, and the necessity of combining results from multiple experiments in order to determine dark matter parameters is discussed.  We demonstrate that spectral differences between the standard dark matter model and a general EFT interaction can produce a bias when calculating exclusion limits and when developing signal models for likelihood and machine learning techniques.  We also discuss the implications of the EFT for the next-generation (G2) direct detection experiments and point out regions of complementarity in the EFT parameter space.

\end{abstract}

\pacs{}
\maketitle


\section{Introduction}

Astrophysical and cosmological evidence indicates that the majority of the matter in the universe takes the form of non-luminous particles called dark matter, though the exact nature of the dark matter particle remains unknown \cite{doi:10.1146/annurev-astro-082708-101659}.  A generic weakly-interacting massive particle (WIMP) is a very attractive dark matter candidate \cite{1985NuPhB.253..375S}. Numerous experiments are engaged in efforts to detect rare collisions between WIMPs and target nuclei in terrestrial detectors.  Results from DAMA \cite{Bernabei:2010uq}, CoGeNT \cite{PhysRevD.88.012002}, CRESST-II \cite{Angloher:2012fk}, and CDMS\,II Si \cite{Agnese:2013rvf} can be interpreted in terms of interactions of WIMPs with masses of 6-30~\gev.   A similar range of masses could also account for a possible excess in the gamma-ray flux near the galactic center in Fermi-LAT data \cite{PhysRevD.84.123005,2014arXiv1409.0042C}.  Under standard assumptions for spin-independent WIMP-nucleon interactions, however, such interpretations are difficult to reconcile with the limits set by CDMSlite \cite{cdmslitePRL}, SuperCDMS \cite{R133_LT}, LUX \cite{PhysRevLett.112.091303}, and PICO \cite{2015arXiv150300008A}.

Standard WIMP scattering calculations make simplifying assumptions about the type of interaction between the nucleon and the dark matter particle: typically only isospin-conserving spin-independent couplings, or spin-dependent couplings to either the proton or neutron are considered. This results in constraints on the three corresponding WIMP-nucleon cross sections.  Relaxing such assumptions can suppress the interaction for some target elements by orders of magnitude relative to others \cite{2013PhRvD..88a5021F}.  In particular, assuming different spin-independent dark matter couplings to protons, $f_p$, and neutrons, $f_n$, can reconcile much of the tension between the CDMS\,II Si allowed region and the SuperCDMS Soudan and LUX exclusion limits \cite{2014JHEP...05..086H}.  However, such solutions often require a high degree of fine-tuning.

In addition, the calculation of dark matter scattering rates typically assumes a Maxwellian velocity distribution \cite{Lewin199687}.  As shown in \cite{2010JCAP...02..030K, 2012JCAP...08..027P}, N-body simulations are not well described by such a distribution.  Consequently, alternate halo models have been proposed.  One such velocity distribution is discussed in \cite{2013ApJ...764...35M,2014PhRvD..89f3513M} and takes the form
\begin{equation}
f(v) = \exp \left [ -\frac{v}{v_0} \right ] \left ( v_{esc}^2 - v^2 \right ) ^p,
\label{eqn:alt_vdist}
\end{equation}
for dark matter velocities smaller than the galactic escape velocity $v_{esc}$.  For values of $v_0/v_{esc}$ and $p$ consistent with N-body simulations, this function falls off faster than the standard Maxwellian distribution.  This difference can significantly affect the expected dark matter event rate, especially for low-mass WIMPs for which experiments are only sensitive to the high-velocity tail of the distribution.  It has been shown that choosing certain values for the parameters of this alternate halo model can reconcile the tension between CDMS\,II Si and XENON100 \cite{2012PhRvL.109r1301A}, though it cannot also account for the tension with LUX because of that experiment's lower energy threshold.

Recently, an effective field theory (EFT) approach for WIMP scattering has been developed that considers all leading-order and next-to-leading order operators  that can occur in the effective Lagrangian that describes the WIMP-nucleus interaction \cite{Fitzpatrick:2012,Anand:2013,Fitzpatrick:2013}.  This formalism introduces new operators that rely on a range of nuclear properties in addition to the standard spin-independent and spin-dependent cases.  It also explicitly includes isospin interference and interference between operators, creating a rich parameter space of possible dark matter interactions that are very sensitive to the specific choice of detector material.

The EFT framework parametrizes the WIMP-nucleus interaction in terms of fourteen operators, $\mathcal O_i$, which are listed in Eq.~\ref{eqn:operators} and include the standard spin-independent and spin-dependent interactions. These operators feature explicit dependence on $\vec v^{\bot}$ (the relative velocity between the incoming WIMP and the nucleon) and the momentum transfer $\vec q$, in addition to the WIMP and nucleon spins, $\vec S_{\chi}$ and $\vec S_N$. Note that $\mathcal O_2$ is not considered since it cannot arise from the non-relativistic limit of a relativistic operator at leading order. In addition, each operator can independently couple to protons or neutrons.  We formulate this isospin dependence in terms of isoscalar and isovector interactions, following the conventions of \cite{Anand:2013}.  

\begin{eqnarray}
\label{eqn:operators}
\nonumber
\mathcal O_1 &=& 1_{\chi} 1_N \\
\nonumber
\mathcal O_3 &=& i \vec S_N \cdot \left [ \frac{ \vec q}{m_N} \times \vec v^{\bot} \right] \\
\nonumber
\mathcal O_4 &=& \vec S_{\chi} \cdot \vec S_N \\
\nonumber
\mathcal O_5 &=& i \vec S_{\chi} \cdot \left [ \frac{ \vec q}{m_N} \times \vec v^{\bot} \right] \\
\nonumber
\mathcal O_6 &=& \left [ \vec S_{\chi} \cdot  \frac{ \vec q}{m_N} \right ]  \left [ \vec S_N \cdot  \frac{ \vec q}{m_N} \right ] \\
\nonumber
\mathcal O_7 &=& \vec S_N \cdot \vec v^{\bot} \\
\nonumber
\mathcal O_8 &=& \vec S_{\chi} \cdot \vec v^{\bot} \\
\nonumber
\mathcal O_9 &=& i \vec S_{\chi} \cdot \left [ \vec S_N \times  \frac{ \vec q}{m_N} \right ] \\
\nonumber
\mathcal O_{10} &=& i \vec S_N \cdot \frac{ \vec q}{m_N} \\
\nonumber
\mathcal O_{11} &=& i \vec S_{\chi} \cdot \frac{ \vec q}{m_N} \\
\nonumber
\mathcal O_{12} &=& \vec S_{\chi} \cdot \left [ \vec S_N \times \vec v^{\bot} \right ] \\
\nonumber
\mathcal O_{13} &=& i \left [ \vec S_{\chi} \cdot \vec v^{\bot}  \right ]  \left [ \vec S_N \cdot  \frac{ \vec q}{m_N} \right ] \\
\nonumber
\mathcal O_{14} &=& i \left [ \vec S_{\chi} \cdot  \frac{ \vec q}{m_N} \right ]  \left [ \vec S_N \cdot   \vec v^{\bot} \right ] \\
\mathcal O_{15} &=& - \left [ \vec S_{\chi} \cdot  \frac{ \vec q}{m_N} \right ] \left [\left (  \vec S_N \times  \vec v^{\bot} \right ) \cdot \frac{ \vec q}{m_N} \right ] 
\end{eqnarray}

These operators contribute to six types of nuclear response functions.  The spin-independent response is denoted by $M$ and is typically the strongest of the six functions since it is related to the number of nucleons in the target nucleus.  The main contribution to this response comes from the standard spin-independent operator $\mathcal O_1$, but it also contains higher-order contributions from operators $\mathcal O_5$, $\mathcal O_8$, and $\mathcal O_{11}$.  There are two spin-dependent responses, $\Sigma'$ and $\Sigma''$, that correspond to projections of spin parallel and perpendicular to the momentum transfer.  A linear combination of these two responses yields $\mathcal O_4$, which is related to the standard spin-dependent response.  Many of the other operators also appear in one of these two responses.  A novel type of response introduced in the EFT, $\Delta$, is related to the net angular momentum of an unpaired nucleon and contains contributions from operators $\mathcal O_5$ and $\mathcal O_8$.  A second novel response is $\Phi''$, which is sensitive to the product of angular momentum and spin.  This response tends to favor heavier elements, and the most dominant contribution to this response is from $\mathcal O_3$. The last response considered in the EFT, $\tilde \Phi '$, contains contributions from operators $\mathcal O_3$, $\mathcal O_{12}$, and $\mathcal O_{15}$.  $\tilde \Phi '$ is discussed less frequently in the literature since it is difficult to find a model that produces this response, but we consider it here for completeness.

The EFT also includes two operator-operator interference terms: $\Sigma' \Delta$ and $M \Phi''$.  $\Sigma'$ interferes with $\Delta$ because velocity-dependent responses are sensitive to properties such as angular momentum that depend on the motion of the nucleon within the nucleus.  This interference term is particularly significant for germanium, which has large responses to both $\Sigma'$ and $\Delta$.  The $\Sigma'\Delta$ response contains interference between $\mathcal O_4$ and $\mathcal O_5$, as well as between $\mathcal O_8$ and $\mathcal O_9$.  In addition, since both $M$ and $\Phi''$ are scalar responses, interference between the two can be significant, especially for elements like xenon that have large responses to both.  The $M \Phi''$ response contains interference between operators $\mathcal O_1$ and $\mathcal O_3$, operators $\mathcal O_{11}$ and $\mathcal O_{12}$, and operators $\mathcal O_{11}$ and $\mathcal O_{15}$.

Since the various responses are related to different nuclear properties, the strength of the resulting interaction can vary by many orders of magnitude.  The expectation values of these properties are listed in \cite{Fitzpatrick:2012}.  For instance, the spin-dependent responses $\Sigma'$ and $\Sigma''$ depend on the square of the spin of an unpaired nucleon, which ranges from $5\times 10^{-6}$ for protons in germanium (which has one isotope with an unpaired nucleon, which is a neutron) to 0.2 for protons in fluorine (which has an unpaired proton).  The angular momentum of a nucleon, which governs the strength of the $\Delta$ response, ranges from $\mathcal O(1\times 10^{-3})$ to $\mathcal O(1)$, while $(L \cdot S)^2$, which governs the strength of the $\Phi''$ response, ranges from 0.1 for light nuclei to several hundred for heavier nuclei.  The strongest response is $M$, which is related to the square of the number of nucleons.

The strength of an EFT interaction is parametrized by numerical coefficients, $c_i^{\tau}$, associated with each operator $\mathcal O_i$, where $\tau=0$ or $1$ denotes the isoscalar ($c_i^0 = \nicefrac{1}{2}( c_i^p + c_i^n$)) and isovector ($c_i^1 = \nicefrac{1}{2}(  c_i^p - c_i^n$)) combinations, respectively.  The coefficients have dimensions of 1/energy$^2$, so we multiply by the weak mass scale ($m_{weak} = 246.2$ GeV) to produce dimensionless quantities.  The $c_i^{\tau}$ are related by a change of basis to generalized versions of $f_n$ and $f_p$ and can take on any value, positive or negative.  The coefficients appear as $c_i^{\tau} c_j^{\tau'}$ in the interaction, indicating that operators interfere pair-wise, at most.  

This paper discusses the EFT approach in the context of current and proposed direct detection experiments.  We present exclusion limits on EFT operator coefficients using the optimum interval method \cite{ upper,2007arXiv0709.2701Y}.  We discuss the differences in energy spectra that arise for arbitrary EFT interactions and examine how this energy dependence may affect future experiments if WIMP candidate events are observed.  We also consider the variation in interaction strength across the elements commonly used as direct detection targets and discuss possible ways of exploring interference using experimental results.  Finally, we discuss the implications of this effective field theory for the next-generation (G2) direct detection experiments, SuperCDMS SNOLAB and LZ.


\begin{figure*}[tb!]
\begin{center}
\includegraphics[height=175 pt]{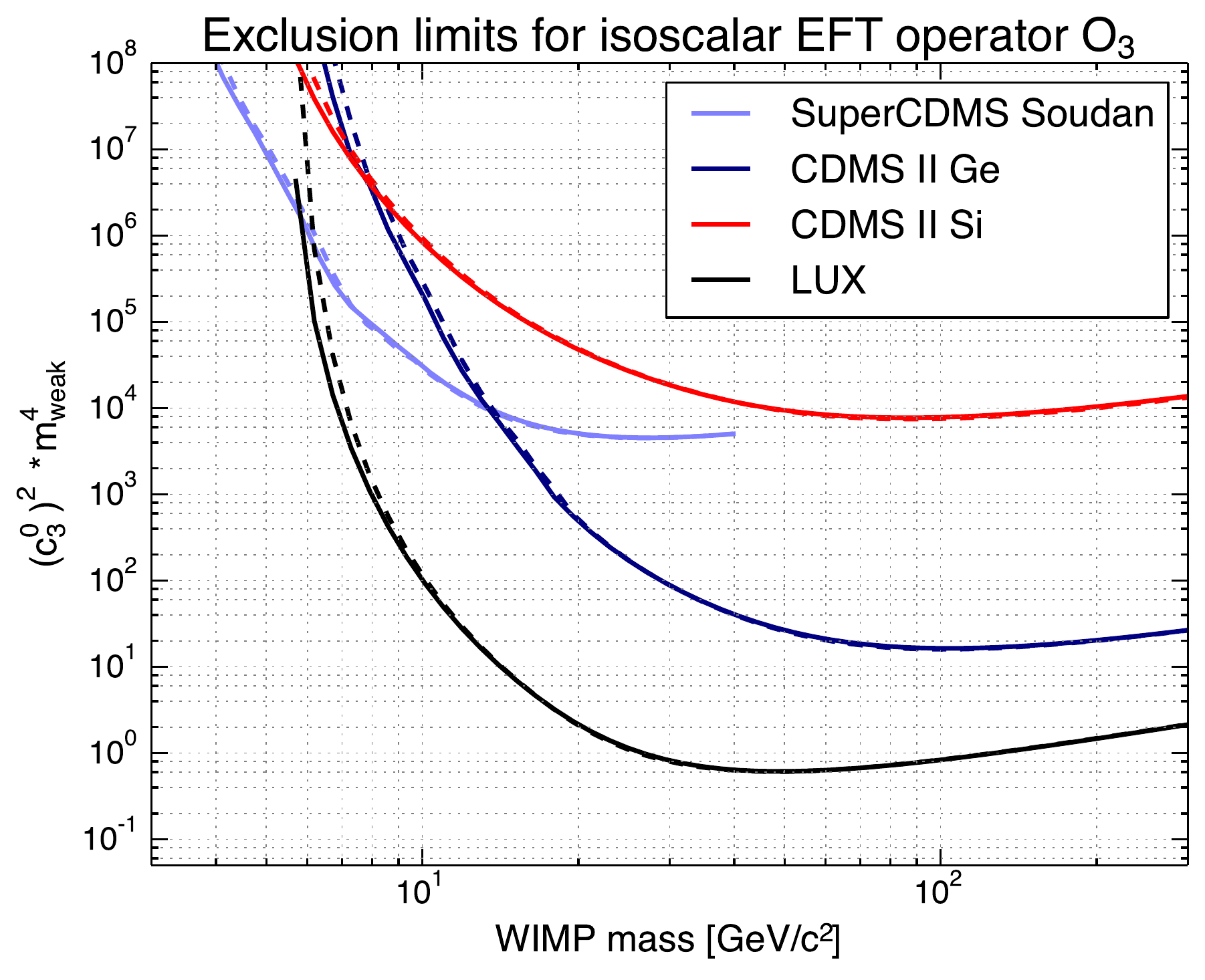}
\includegraphics[height=175 pt]{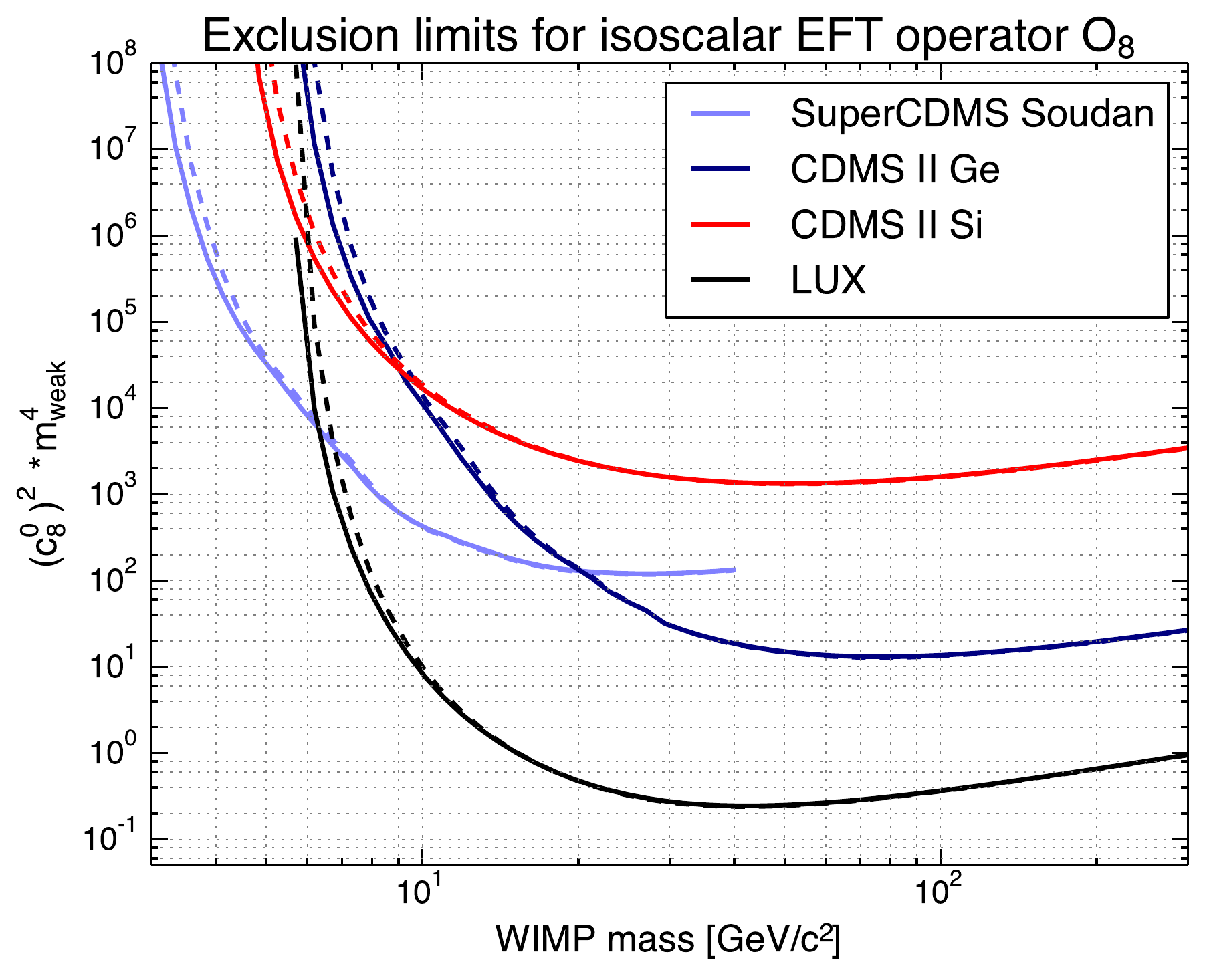}
\caption{Upper limits on the dimensionless isoscalar coefficients $c_3^0$ (left) and $c_8^0$ (right) as a function of WIMP mass for SuperCDMS Soudan (light blue) \cite{R133_LT}, CDMS\,II Ge reanalysis (dark blue) \cite{CDMS-II_reanalysis}, and CDMS\,II Si (red) \cite{c34_Si}, and estimated limits for LUX (black) \cite{PhysRevLett.112.091303}, for the Maxwellian halo (solid) and an alternate halo model (dashed).  }
\label{fig:O3_O8_limits}
\end{center}
\end{figure*}

\section{Exclusion limits on a set of EFT operators}

The strength of the interaction in the EFT framework is governed by a set of 28 numerical coefficients corresponding to the 14 operators, one for each isospin.  Others have attempted to find global fits in this multi-dimensional parameter space, combining data from many direct detection experiments \cite{2014JCAP...09..045C}.  Since the parameter space is large and relatively unconstrained by current experiments, we choose instead to calculate exclusion limits on the coefficients for individual EFT operators for three different target elements: germanium (SuperCDMS Soudan and CDMS\,II), silicon (CDMS\,II), and xenon (LUX).  This paper presents the first EFT experimental result that includes all three target elements that will be used in the G2 experiments.

We use the optimum interval method to calculate 90\% upper confidence limits on the numerical coefficients of EFT operators. The optimum interval method incorporates information about the candidate event energies and energy-dependent detection efficiencies, which can yield stronger exclusion limits in the presence of unknown backgrounds than likelihood methods that consider only a single energy bin in the presence of backgrounds. This is particularly important here because of the spectral differences that can arise from different EFT interactions.  We consider a single operator at a time and present the exclusion limit on the square of the EFT coefficient, which is proportional to the total interaction cross section.  We compare the effects of two halo models on the limits. The first uses standard halo assumptions as in \cite{Lewin199687}, with a WIMP mass density $\rho_0$ = 0.3 GeV/$c^2$/cm$^3$, most probable WIMP velocity of 220\,km/s, mean circular velocity of the Earth with respect to the galactic center of 232\,km/s, galactic escape velocity of 544\,km/s, and a velocity distribution that correctly takes into account the effect of the Earth's velocity on the escape-velocity cutoff \cite{arrenberg}.  The second halo model uses the functional form of Eq.~\ref{eqn:alt_vdist} with $p=2.7$ and $v_0/v_{esc} = 0.6875$, determined by fits to the Eris simulation of a Milky-Way-like galaxy \cite{Eris_DMdistribution}, and other halo parameters as above.

Figure\,\ref{fig:O3_O8_limits} shows the upper limits for two example operators, isoscalar operators ${\cal O}_3$ (left) and ${\cal O}_8$ (right), as a function of WIMP mass.  Limits on all operators for a small range of masses can be found in Table \ref{limits_table}. Limits on all operators for a small range of masses can be found in Table \ref{limits_table}.  Solid lines correspond to the Maxwellian halo, whereas dashed lines show the limit calculated assuming the alternate velocity distribution function discussed above.  The SuperCDMS Soudan, CDMS\,II Ge (reanalysis), and CDMS\,II Si limits use the candidate events, thresholds, and detection efficiencies discussed in \cite{R133_LT}, \cite{CDMS-II_reanalysis}, and \cite{c34_Si} respectively, while the estimated LUX limit assumes zero observed events and functional form for the detection efficiency that follows a hyperbolic tangent versus energy centered at 2.5\,keV$_{\mbox{nr}}$ but with a step function cutoff that goes to zero below 3\,keV$_{\mbox{nr}}$. 

\begin{table*}[ht]
\centering 
\begin{tabular}{c@{\hskip 12pt} c@{\hskip 15pt} c@{\hskip 15pt} c} 
\hline\hline    
Operator coefficient   & SuperCDMS Soudan              &  CDMS\,II Ge	& CDMS\,II Si\\     \hline             
$(c^0_{1})^2 * m_{weak}^4$ & $8.98 \times 10^{-5}$ (---)             & $2.00 \times 10^{-3}$ ($8.42 \times 10^{-6}$) & $3.06 \times 10^{-3}$ ($7.73 \times 10^{-4}$)  \\
$(c^0_{3})^2 * m_{weak}^4$& $3.14 \times 10^{4}$ (---)             & $2.24 \times 10^{5}$ ($ 2.66 \times 10^{1}$) & $8.59 \times 10^{5}$ ($1.37 \times 10^{4}$)  \\
$(c^0_{4})^2 * m_{weak}^4$& $8.77 \times 10^{1}$ (---)             & $2.05 \times 10^{3}$ ($1.10 \times 10^{1}$) & $3.94 \times 10^{3}$ ($1.02 \times 10^{3}$)  \\
$(c^0_{5})^2 * m_{weak}^4$ & $6.34 \times 10^{5}$ (---)             & $9.18 \times 10^{6}$ ($4.04 \times 10^{3}$) & $2.67 \times 10^{7}$ ($1.55 \times 10^{6}$)  \\
$(c^0_{6})^2 * m_{weak}^4$ & $4.54 \times 10^{8}$ (---)             & $3.30 \times 10^{9}$ ($4.50 \times 10^{5}$) & $2.44 \times 10^{10}$ ($3.70 \times 10^{8}$)  \\
$(c^0_{7})^2 * m_{weak}^4$ & $8.44 \times 10^{7}$ (---)             & $2.51 \times 10^{9}$ ($1.12 \times 10^{7}$) & $3.19 \times 10^{9}$ ($929 \times 10^{8}$)  \\
$(c^0_{8})^2 * m_{weak}^4$ & $4.30 \times 10^{2}$ (---)             & $1.16 \times 10^{4}$ ($2.67 \times 10^{1}$) & $1.70 \times 10^{4}$ ($3.49 \times 10^{3}$)  \\
$(c^0_{9})^2 * m_{weak}^4$ & $1.95 \times 10^{5}$ (---)             & $2.48 \times 10^{6}$ ($3.87 \times 10^{3}$) & $9.17 \times 10^{6}$ ($7.34 \times 10^{5}$)  \\
$(c^0_{10})^2 * m_{weak}^4$ & $9.22 \times 10^{4}$ (---)             & $1.11 \times 10^{6}$ ($9.08 \times 10^{2}$) & $4.34 \times 10^{6}$ ($2.86 \times 10^{5}$)  \\
$(c^0_{11})^2 * m_{weak}^4$ & $5.13 \times 10^{-1}$ (---)             & $6.15 \times 10^{0}$ ($5.46 \times 10^{-3}$) & $1.86 \times 10^{1}$ ($1.34 \times 10^{0}$)  \\
$(c^0_{12})^2 * m_{weak}^4$ & $1.03 \times 10^{2}$ (---)             & $1.21 \times 10^{3}$ ($8.70 \times 10^{-1}$) & $2.45 \times 10^{3}$ ($1.69 \times 10^{2}$)  \\
$(c^0_{13})^2 * m_{weak}^4$ & $4.28 \times 10^{8}$ (---)             & $3.06 \times 10^{9}$ ($3.56 \times 10^{5}$) & $2.50 \times 10^{13}$ ($1.36 \times 10^{12}$)  \\
$(c^0_{14})^2 * m_{weak}^4$ & $5.00 \times 10^{11}$ (---)             & $8.20 \times 10^{12}$ ($8.46 \times 10^{9}$) & $2.64 \times 10^{13}$ ($1.72 \times 10^{12}$)  \\
$(c^0_{15})^2 * m_{weak}^4$ & $1.32 \times 10^{8}$ (---)             & $5.65 \times 10^{8}$ ($1.10 \times 10^{4}$) & $4.44 \times 10^{9}$ ($1.48 \times 10^{7}$)  \\

\hline\hline
\end{tabular}
\caption{SuperCDMS  and CDMS\,II 90\% confidence level upper limits on the square of the dimensionless EFT coefficient for pure isoscalar interaction for a 10 \gev~(300 \gev) WIMP for all isoscalar EFT operators.  The upper limits vary in accordance with the relative strength of the interaction in silicon and germanium.}
\label{limits_table}
\end{table*}

Because of the different nuclear responses for the three target elements considered, the relative strength of the limits varies from operator to operator.  In particular, $\mathcal O_8$ (Fig.~\ref{fig:O3_O8_limits}, right) includes contributions from the $\Delta$ response, which is greater in germanium than in silicon or xenon. This contribution strengthens the SuperCDMS Soudan constraint relative to LUX and CDMS\,II Si.  In addition, the shape of the curve for a single target element changes from operator to operator.  For example, $\mathcal O_3$ depends on the square of the momentum transfer, naturally suppressing the event rate at low energies.  As a result, the limits at low WIMP mass for $\mathcal O_3$ are weaker than for other operators.

The difference between the two WIMP velocity distributions becomes apparent when the only events expected above the detection thresholds are due to WIMPs in the high-velocity tails.  Since both CDMS and LUX have thresholds of a few keV, this disparity appears only at the lowest WIMP masses.  The difference is also more pronounced for LUX, since its target nucleus, xenon, is heavier than silicon or germanium.  A dark matter particle must have a higher velocity to deposit a given recoil energy in xenon than in germanium or silicon; higher-energy recoils become comparatively rarer.  For the SuperCDMS Soudan result, the difference in velocity distributions leads to a factor of two difference in the limit around 4\,\gev, whereas for LUX, the difference in velocity distribution leads to a factor of two difference around 7\,\gev.

\begin{figure}[htb!]
\begin{center}
\includegraphics[width=250 pt]{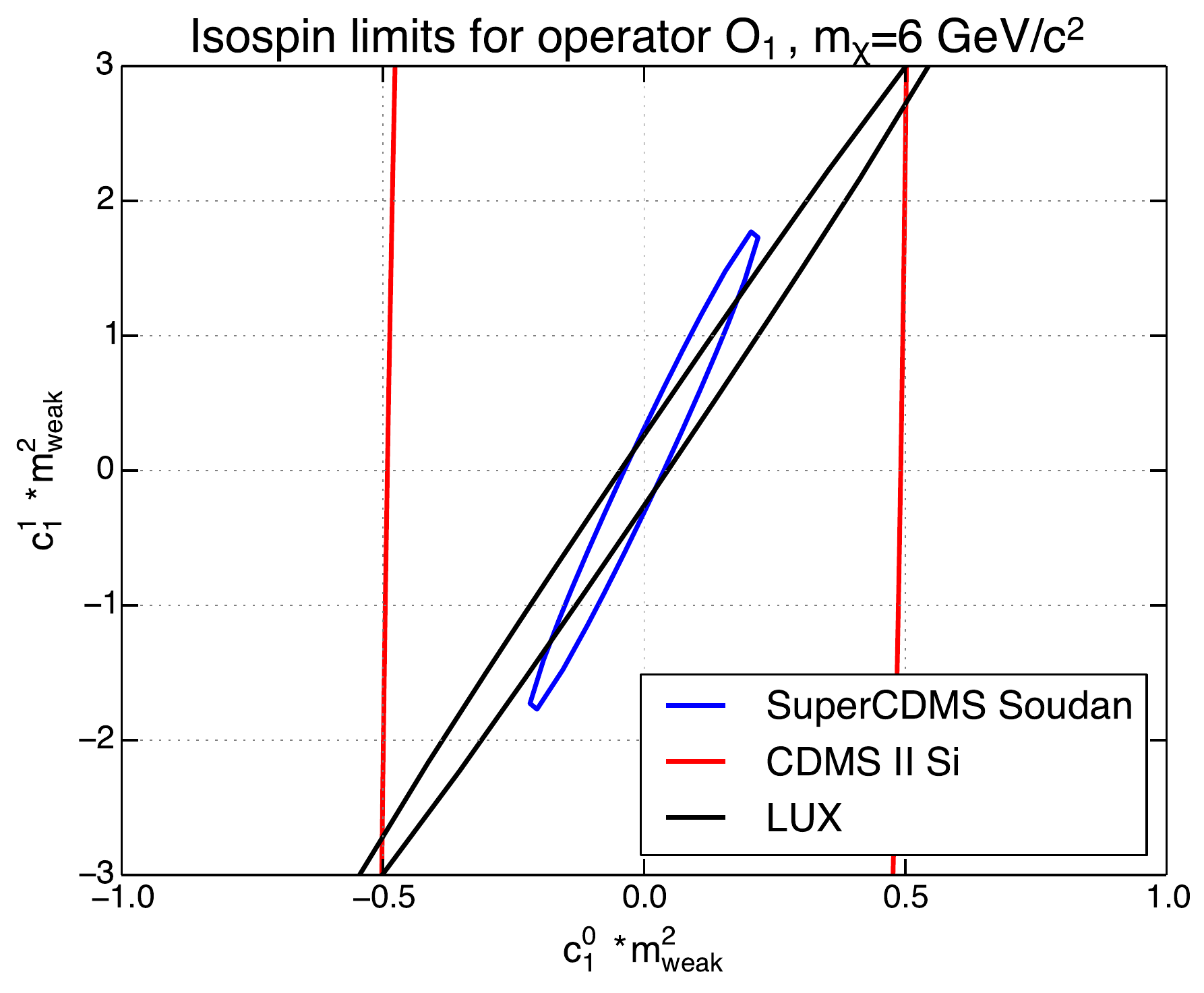}
\caption{Polar limits on $\mathcal O_1$ isospin for SuperCDMS Soudan (blue) \cite{R133_LT}, LUX \cite{PhysRevLett.112.091303} (black), and CDMS II Si (red) \cite{c34_Si} at a WIMP mass of 6\,\gev.   }
\label{fig:O1_isospin}
\end{center}
\end{figure}

Since the EFT explicitly includes isospin dependence, we can also use the optimum interval method to set polar limits on isospin.  For a given WIMP mass and a given angle between the isoscalar and isovector components of an operator, we set a 90\% upper confidence limit on the isoscalar-isovector radius. Varying the polar angle produces exclusion ellipses in the isoscalar-isovector plane, as in Fig.~\ref{fig:O1_isospin}, which shows limits for operator $\mathcal O_1$ and a 6\,\gev~WIMP.  The major axis of each ellipse corresponds to the value of $c_1^0/c_1^1$ that yields maximum suppression of the scattering rate.  Note that although the exposures for CDMS and SuperCDMS are significantly lower than for LUX, there are regions of parameter space allowed by LUX but excluded by SuperCDMS and CDMS at 90\% confidence. This example demonstrates that a combination of experiments using several target nuclei can constrain the EFT parameter space better than any single experiment.


\begin{figure*}[htb!]
\begin{center}
\includegraphics[height=175 pt]{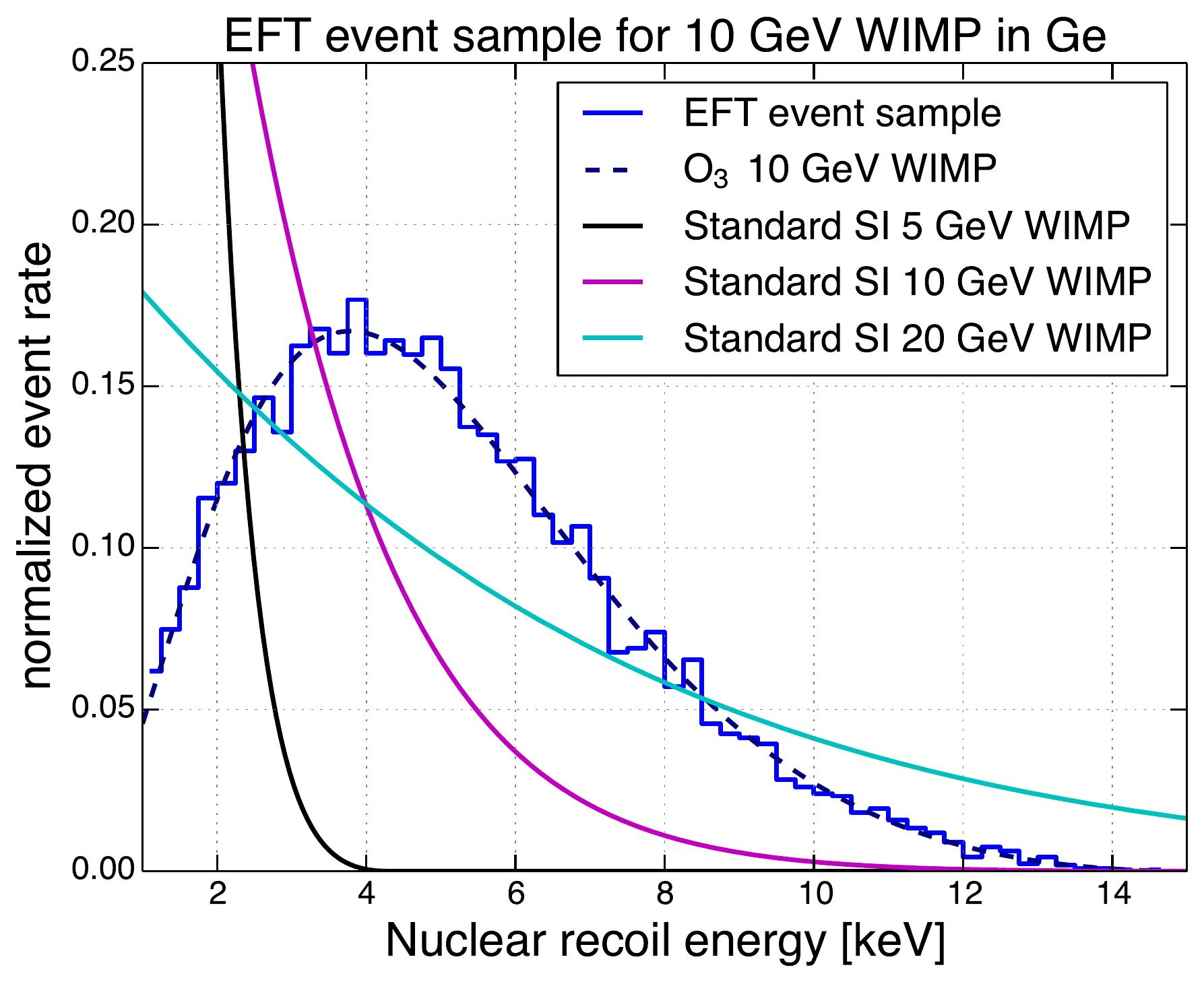}
\includegraphics[height=175 pt]{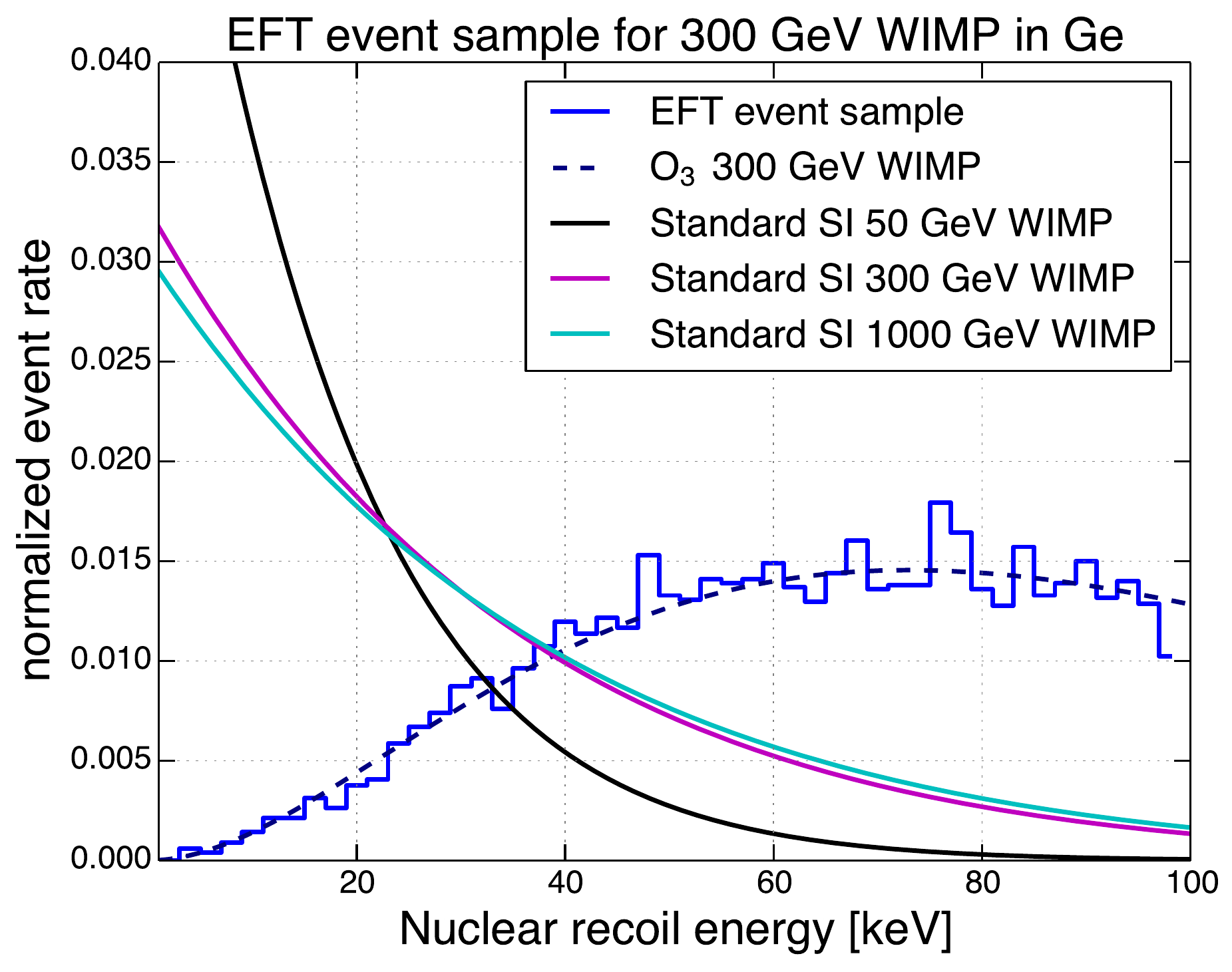}
\caption{Co-added energy spectrum from 100 simulated experiments (blue histogram) assuming the dark matter interaction proceeds according to the isoscalar $\mathcal O_3$ operator for a 10\,\gev~(left) and a 300\,\gev~WIMP (right).  The detection efficiency is assumed to be independent of energy.  The smooth cyan, magenta, and black curves show the expected spectrum for the standard spin-independent rate for several WIMP masses, while the dashed dark blue curve shows the $\mathcal O_3$ spectrum from which the simulated experiments were sampled.}
\label{fig:O3_event_samples}
\end{center}
\end{figure*}

\section{Effect of EFT energy dependence on standard limits}

Because of the additional momentum dependence of several of the EFT operators, the differential event rate for an arbitrary dark matter interaction could be very different than for the standard calculation.  Consequently, it is possible that a limit-setting algorithm that expects the (approximately) exponential event rate of the standard spin-independent interaction could misinterpret a potential signal from a more general EFT interaction as background.

To demonstrate the possible bias that could arise from assuming the standard spin-independent event rate when setting limits, we perform simulated experiments assuming that the dark matter scattering is purely due to a single isoscalar EFT operator.  Figure \ref{fig:O3_event_samples} shows the co-added results of 100 simulated experiments sampled from the energy spectrum of isoscalar $\mathcal O_3$ scattering in germanium for two different dark matter masses, assuming an energy-independent (or ``flat") detection efficiency.  The operator coefficients were set to give each simulated experiment an expectation value of 10 events. This expectation was then convolved with a Poisson distribution to select the number of events for a given simulated experiment.  %

\begin{figure*}[htb!]
\begin{center}
\includegraphics[height=175 pt]{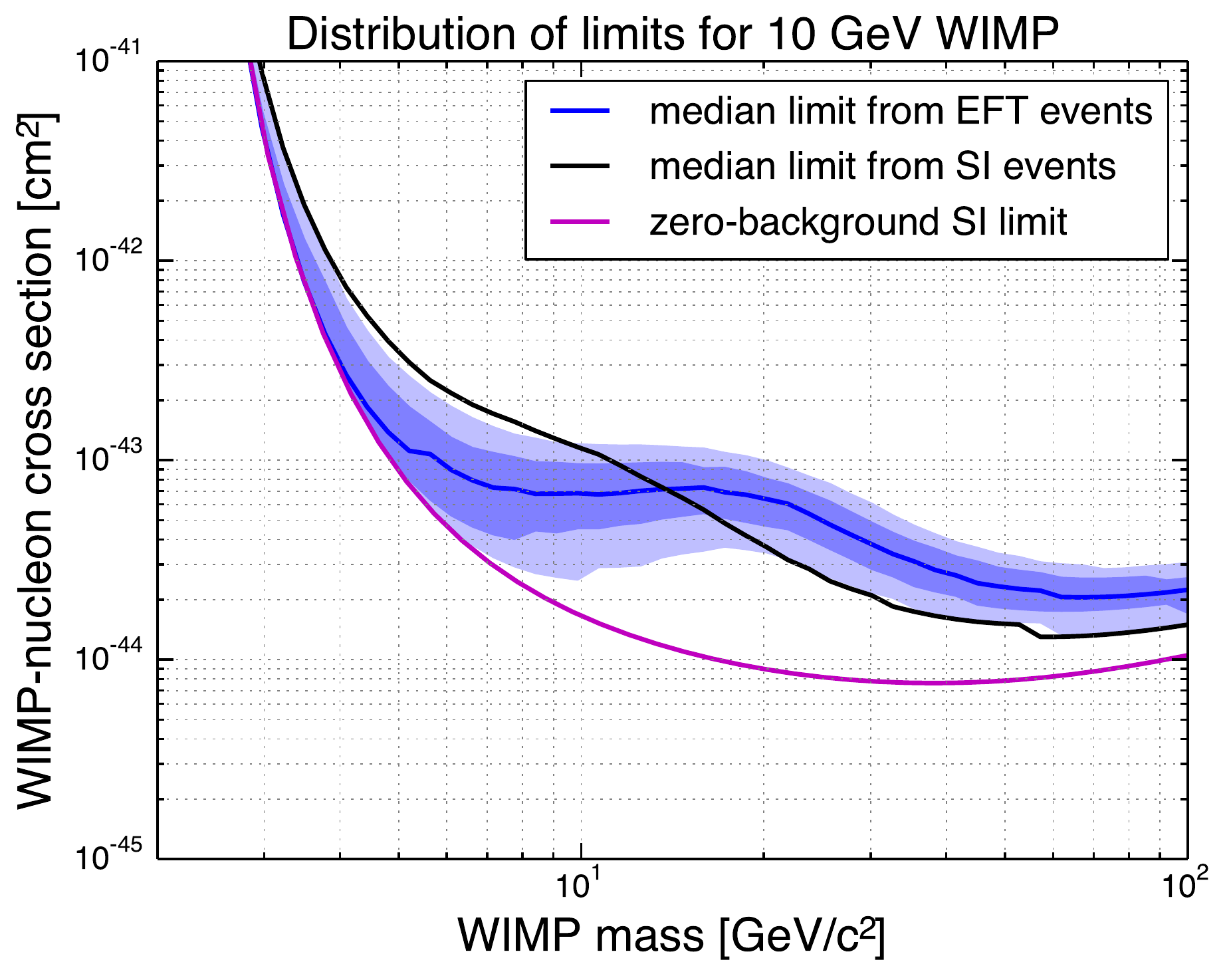}
\includegraphics[height=175 pt]{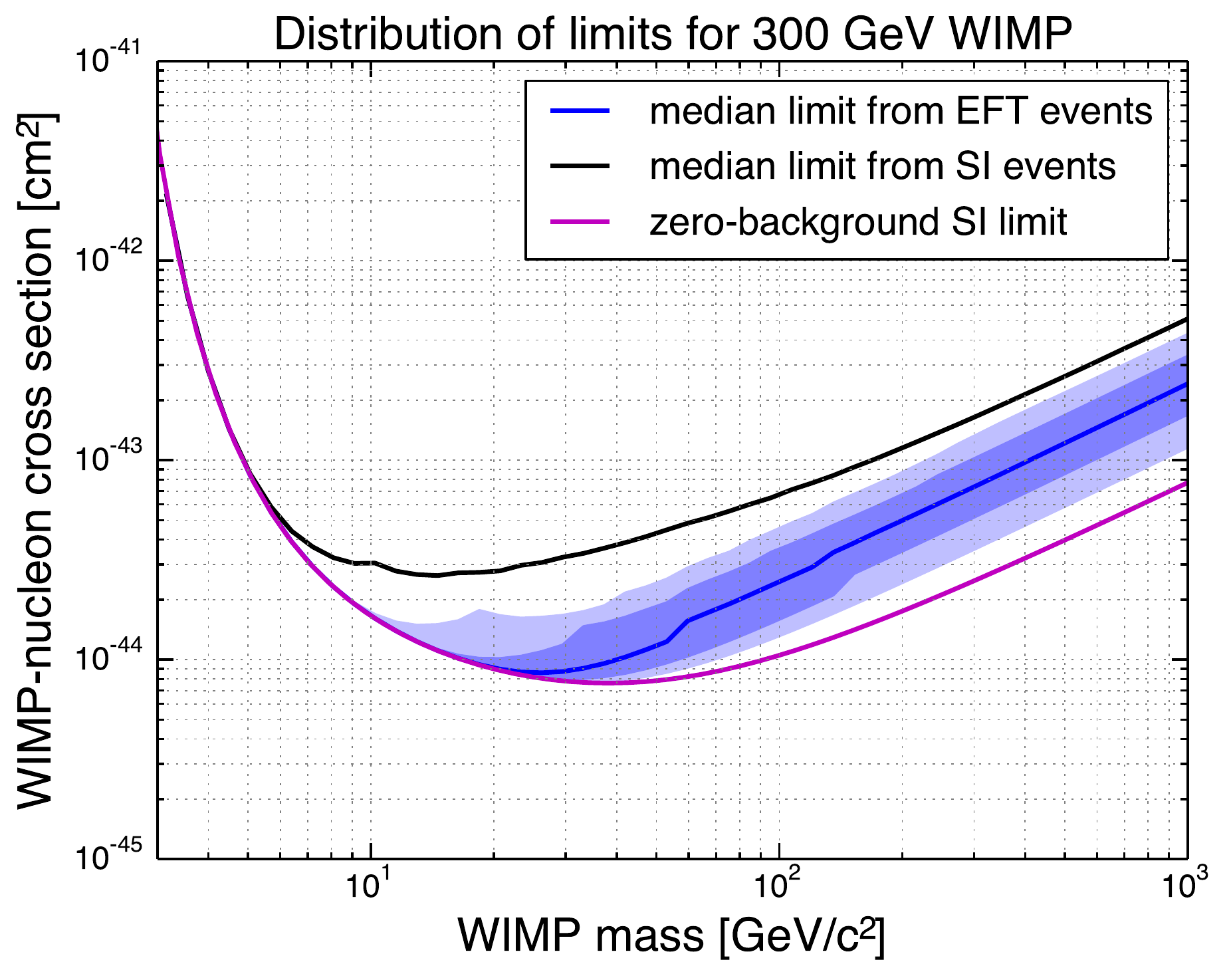}
\caption{Distribution of 90\% confidence level upper limits calculated using the optimum interval method for the simulated experiments discussed in Sec.~3 and shown in Fig.~3, sampled from the event rate for isoscalar $\mathcal O_3$.  Shaded blue bands show the 68\% and 95\% confidence level uncertainty on the distribution.  The zero-background Poisson limit is shown in magenta. }
\label{fig:O3_limits}
\end{center}
\end{figure*}

Unlike the standard spin-independent event rate, the event rate for $\mathcal O_3$ depends on the square of the momentum transfer, so the event rate is suppressed at low recoil energies.  This effect is illustrated in Fig.~\ref{fig:O3_event_samples}, where the black, cyan, and magenta curves show the standard spin-independent scattering rate for a range of WIMP masses and the blue histogram corresponds to the simulated spectrum expected from ${\cal O}_3$ interactions.  For the case of a 10~\gev~WIMP mass, the distribution of events is more closely matched by the spin-independent rate for a higher-mass WIMP. For the 300~\gev~case, no spin-independent rate calculation matches the observed spectrum of events; if experimenters only consider the spin-independent WIMP rate, they may erroneously conclude that they have observed an unexpected background or incorrectly measured their detection efficiency as a function of energy.


We calculate the 90\% confidence level upper limit on the spin-independent cross section for each simulated experiment using the optimum interval method \cite{2007arXiv0709.2701Y, upper} and the standard Maxwellian halo model with halo parameters as above with no background subtraction.  Each simulated experiment was assumed to have an exposure of 1000 kg days and a flat efficiency of 60\% between 1 and 100\,keV$_{\mbox{nr}}$.  The distribution of limits is shown in Fig.~\ref{fig:O3_limits}.  Figure~\ref{fig:O3_limits} also shows the resulting median limit from simulated experiments sampled from the spin-independent distribution in black.

The distribution of limits on the spin-independent cross section for the simulated experiments sampled from the $\mathcal O_3$ energy spectrum deviates from the zero-background limit shown in magenta as well as from the mean limit derived from similar simulated experiments sampling from the spin-independent rate.  As expected, the simulated-experiment limits are weaker than the zero-background limits due to the presence of candidate events.  However, because the energy distribution of the candidate events sampled from $\mathcal O_3$ is different than the expected spin-independent rate, the limits also deviate from the expected shape for the true spin-independent experiment.

In the 10\,\gev~case, we expect the limit to be weakest around a mass of 10\,\gev, where the rate expected by the limit algorithm matches the observed event rate.  However, because the observed events due to $\mathcal O_3$ scattering are skewed towards higher recoil energies, the limit tends to be weaker at larger WIMP masses where the tail of the spin-independent event rate extends to higher recoil energies.  For the 300\,\gev~case, the distribution of limits agrees with the Poisson zero-background limit at low masses; the observed events occur at recoil energies that cannot be produced by a low-mass WIMP.  At higher masses, the distribution of limits is still close to the zero-background limit because the shape of the observed spectrum is very different from the expected spin-independent WIMP rate.

The difference in the limits between the spin-independent and EFT cases demonstrates the importance of correctly modeling the expected WIMP signal.  Algorithms that assume the standard spin-independent rate when calculating limits will interpret events from EFT interactions with different spectral shapes as background, and thus, this assumption could lead to a bias in the exclusion limits reported by experiments, especially in the case where events are observed.  
\section{Interference in the EFT parameter space}

\subsection{General interference framework}

The EFT framework also provides a more general description of interference among operators such as the ``xenophobic" isospin violation case discussed in the literature  \cite{2013PhRvD..88a5021F}.  It not only allows for interference between the isospin components of individual operators, but also among different operators.  The generalized interference can be written as a matrix equation in the large EFT parameter space, but because operators interfere in pairs, and only certain pairs interfere, this large matrix can be decomposed into block-diagonal form.  We consider the $2 \times 2$ case of isospin interference and the $4 \times 4$ case of isospin and operator-operator interference.

The generalized amplitude for the $4 \times 4$ case can be written as the product of the vector of operator coefficients $c_i^{\tau}$ with the amplitude matrix, where superscript 0 and 1 indicate isoscalar and isovector, respectively, and the subscripts indicate the operator being considered:

\begin{equation}
\left[ \begin{array}{cccc}
c^0_i & c^1_i & c^0_j & c^1_j\\
\end{array} \right ]
\left[ \begin{array}{cccc}
A^{00}_{ii} & A^{01}_{ii} & A^{00}_{ij} & A^{01}_{ij}\\
A^{10}_{ii} & A^{11}_{ii} & A^{10}_{ij} & A^{11}_{ij}\\
A^{00}_{ji} & A^{01}_{ji} & A^{00}_{jj} & A^{01}_{jj}\\
A^{10}_{ji} & A^{11}_{ji} & A^{10}_{jj} & A^{11}_{jj}
\end{array} \right]
\left [ \begin{array}{c}
c^0_i \\
c^1_i\\
c^0_j\\
c^1_j
\end{array}\right].
\label{eqn:matrix}
\end{equation}
The amplitudes $A^{\tau \tau'}_{ij}$ are the product of the WIMP and nuclear response functions for the interaction specified by $c^{\tau}_i$ and $c_j^{\tau'}$ and depend on properties such as target element, WIMP mass, WIMP spin, WIMP velocity, and nuclear recoil energy.  We evaluate the $A^{\tau \tau'}_{ij}$ without integrating over the dark matter velocity distribution to avoid introducing more variables.  Amplitudes are summed over the isotopes for a given element according to their natural abundances.

Finding the eigenvectors of this matrix will give the ``principal components" of the interaction space.  We expect that three of the four eigenvalues should be small, since the matrix for a single isotope is an outer product and therefore should have a single nonzero eigenvalue.  The vector with the largest eigenvalue corresponds to the maximal amplitude for scattering in the interference space under consideration, while the three small eigenvalues correspond to local extrema in the scattering amplitude which tend to suppress the event rate.  To be maximally sensitive to the parameter space for a given interference case, we would like to choose target elements whose constructive interference eigenvectors span the space of interactions.

\begin{figure}[htb]
\begin{center}
\includegraphics[width=\columnwidth]{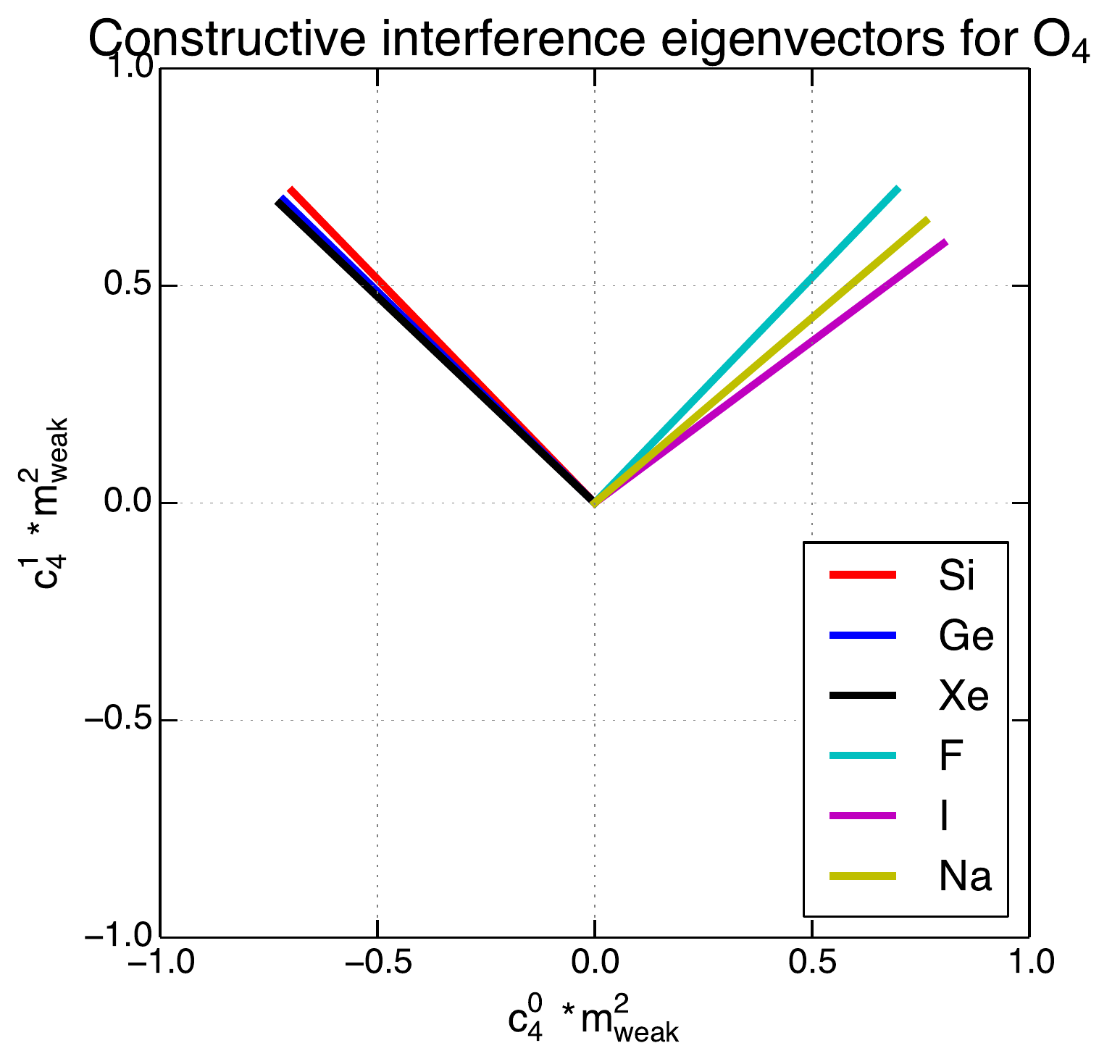}
\caption{Constructive interference eigenvectors for 2D $\mathcal O_4$ isospin interference.  Proton-dominated interactions occur along the $x=y$ diagonal, while neutron-dominated interactions occur along the $x=-y$ diagonal.}
\label{fig:O4_isospin}
\end{center}
\end{figure}

As an example, we first consider isospin interference for a single operator in an already well-understood case.  Figure \ref{fig:O4_isospin} shows the constructive isospin interference eigenvectors for scattering via operator $\mathcal O_4$ (the standard spin-dependent operator) for several elemental targets, evaluated at a WIMP mass of 100\,\gev~and nuclear recoil energy of 100 keV.  The vectors are plotted in the space of the isoscalar coefficient versus the isovector coefficient.  The proton-neutron space can be recovered from this basis via a 45-degree rotation.  The amplitude in a given direction indicates the target's response to that operator and illustrates the sensitivity of each material to the corresponding operator.  In addition, if we were to plot polar limits as in Fig.~\ref{fig:O1_isospin} for $\mathcal O_4$, we would see that the direction of the constructive interference vector corresponds to the minor axis of the ellipse.
 In the two-dimensional case, the destructive interference vector is perpendicular to the constructive vector and corresponds to the major axis of the ellipse in a polar limit plot.

Since $\mathcal O_4$ is the standard spin-dependent operator, we see that the constructive interference eigenvectors fall into two categories based on the nucleon content of the target nucleus.  The elements with unpaired protons (fluorine, sodium, and iodine) have maximal scattering rates when the interaction is proton-dominated, corresponding to $c^0 = c^1$.  On the other hand, the elements with unpaired neutrons (germinum, xenon, and silicon) have maximal scattering rates when the interaction is neutron-dominated, corresponding to $c^0 = -c^1$.  Consequently, to span this space and therefore be maximally sensitive to all possible spin-dependent interactions, we should choose one element each from the neutron- and proton-dominated sets.

\begin{figure}[htb]
\begin{center}
\includegraphics[width=\columnwidth]{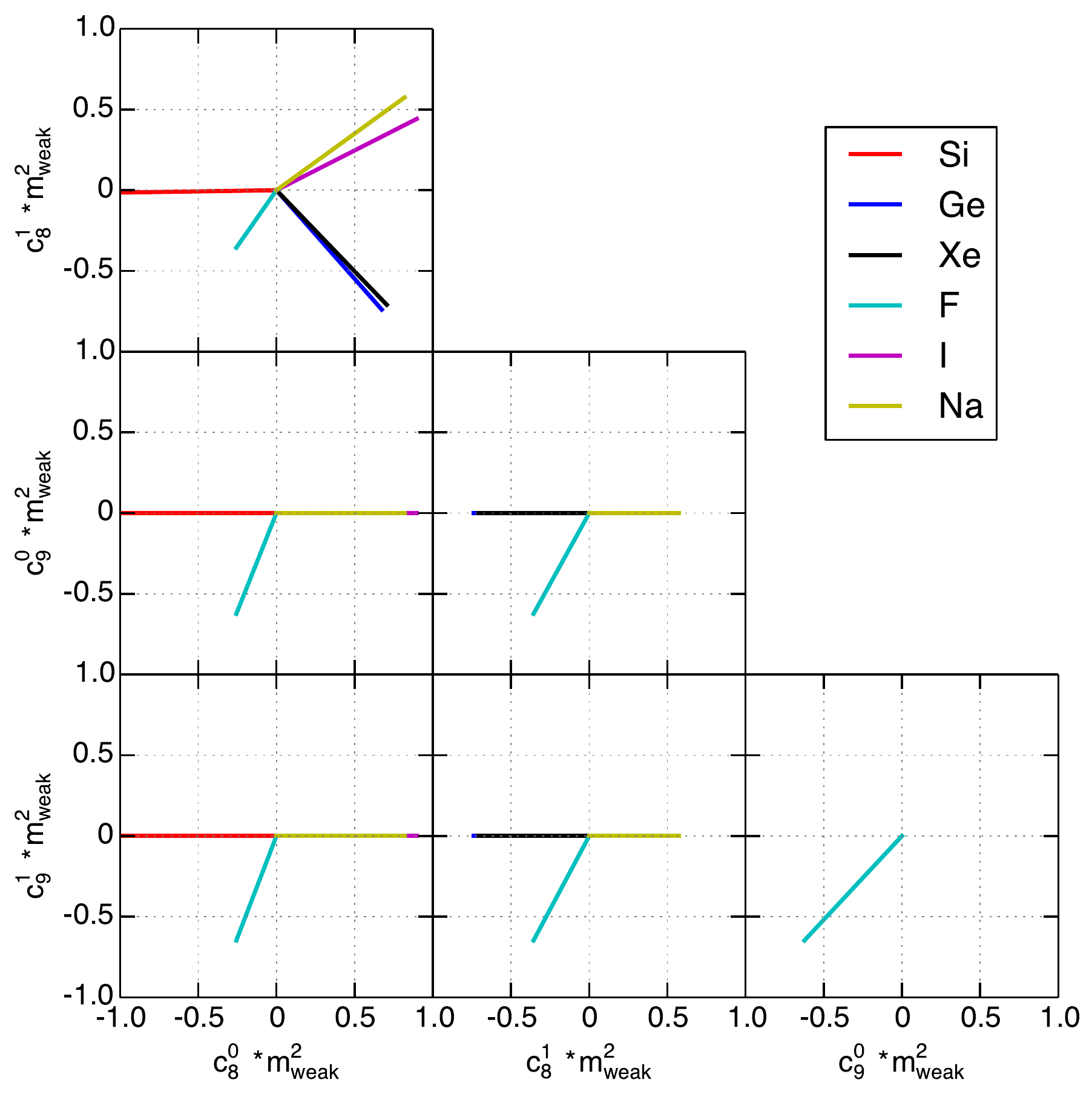}
\caption{Constructive interference eigenvectors for 4D $\mathcal O_{8} / \mathcal O_{9}$ interference.  }
\label{fig:O8_O9_constructive}
\end{center}
\end{figure}

We can apply this same procedure to the more general 4D case to demonstrate the complementarity of the different target elements.  As an example, Figure \ref{fig:O8_O9_constructive} shows all 2D projections of the four-dimensional eigenvectors in the interference space for $\mathcal O_{8}$ and  $\mathcal O_{9}$, evaluated for a WIMP mass of 100\,\gev~and nuclear recoil energy of 30 keV.  The eigenvectors for scattering in silicon, germanium, xenon, iodine, and sodium indicate that they are most sensitive to various combinations of isoscalar and isovector $\mathcal O_{8}$ scattering.  However, the vector for fluorine shows that it is sensitive to both $\mathcal O_{8}$ and  $\mathcal O_{9}$.  This variation across targets allows different experiments to probe different regions of the EFT parameter space, increasing the overall sensitivity of the direct detection method.

\begin{figure}[htb!]
\begin{center}
\includegraphics[width=0.92\columnwidth]{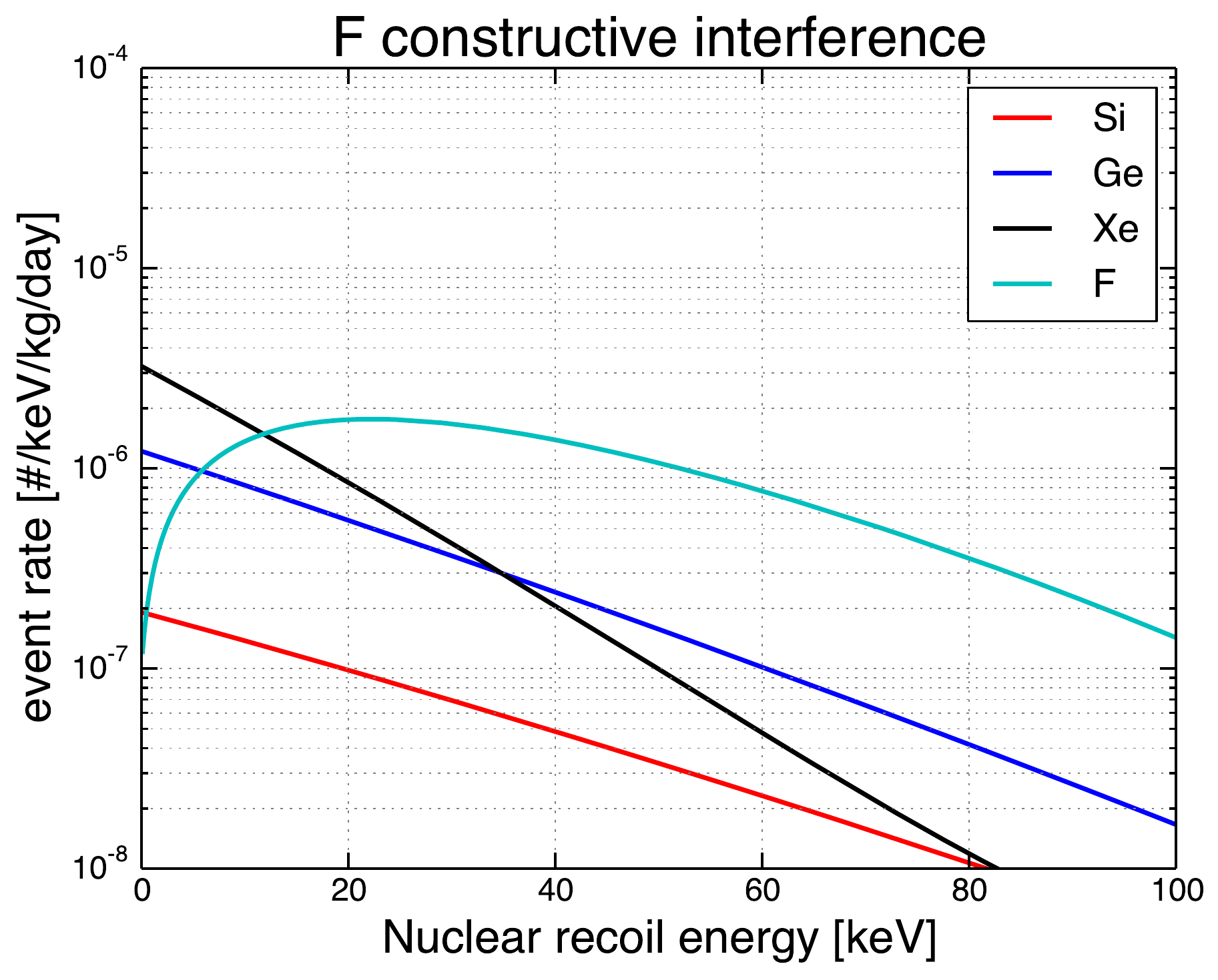}
\includegraphics[width=0.92\columnwidth]{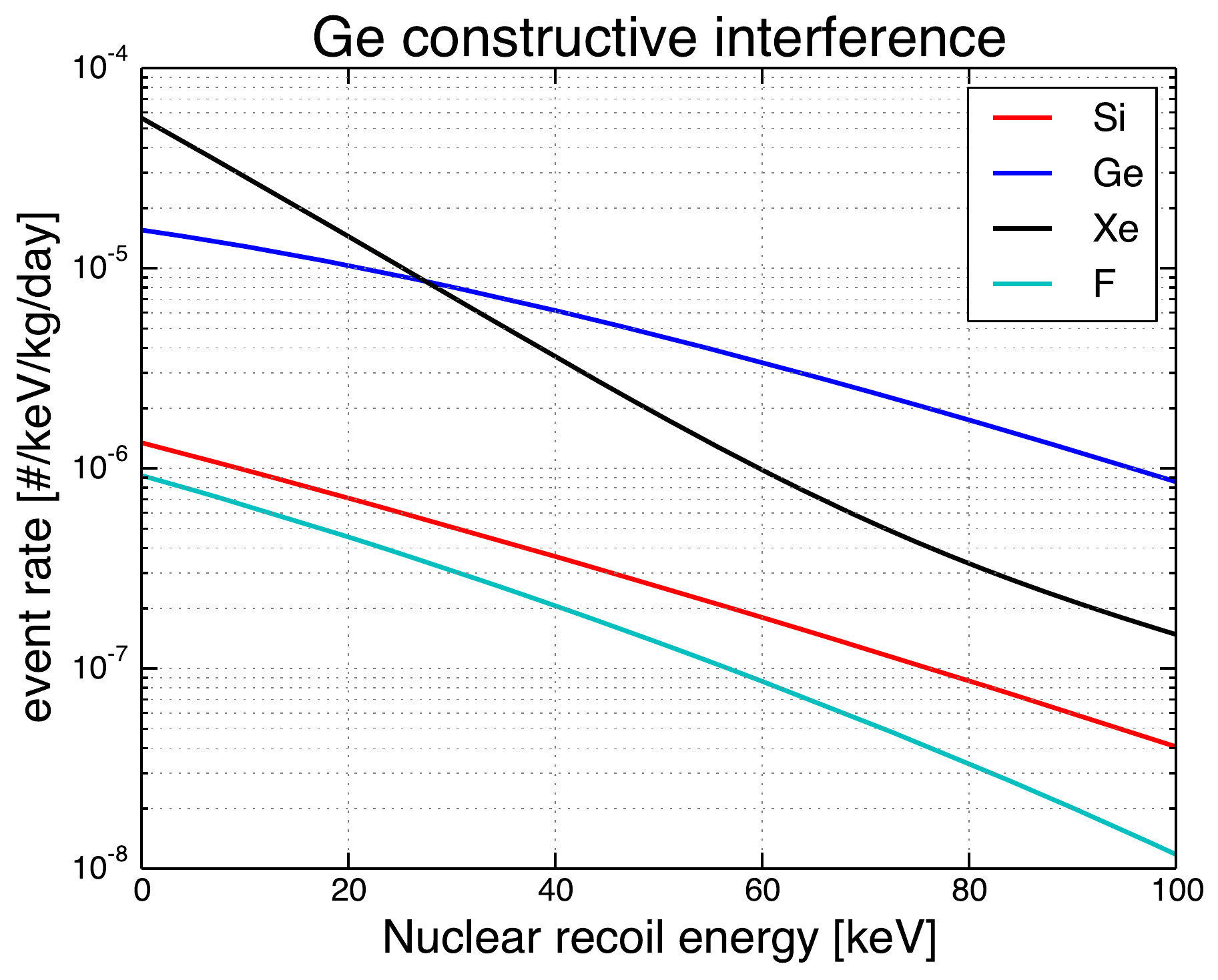}
\caption{Differential event rate evaluated at the $\mathcal O_{8}/\mathcal O_{9}$ constructive interference vector from Fig.~\ref{fig:O8_O9_constructive} for fluorine (top) and germanium (bottom).}
\label{fig:interference_evt_rate}
\end{center}
\end{figure}

To demonstrate the effect of this four-dimensional interference on the differential event rate, we evaluate the event rate using the operator coefficients from two four-dimensional interference eigenvectors from Fig.~\ref{fig:O8_O9_constructive} that point in different directions in the parameter space.  Figure \ref{fig:interference_evt_rate} shows the differential event rate for several targets evaluated at the constructive interference vectors for fluorine (top) and germanium (bottom) for $\mathcal O_{8}/\mathcal O_{9}$ interference.  Since the fluorine eigenvector is not parallel to the germanium eigenvector, the germanium event rate evaluated at the fluorine vector is suppressed and vice versa.  In addition, since the xenon and germanium eigenvectors are nearly parallel in this case, the two event rates are comparable at the 30\,keV nuclear recoil energy at which the eigenvectors are evaluated.

This example shows the large variation in signal strength that can occur for different combinations of operators.  In this case, varying the coefficients from the germanium eigenvector to the fluorine eigenvector led to an order-of-magnitude suppression of the rate in germanium, silicon, and xenon, and a change in the energy spectrum for fluorine.  Similar suppression can also occur for the other interference terms present in the effective field theory.

\subsection{Comparison of target elements for G2 direct detection experiments}

Three target elements will be used in the upcoming G2 experiments:  germanium, silicon, and xenon.  Under the standard spin-independent scattering framework, where the rate scales as $\sim$$A^2$, experiments that use xenon as a target element have the greatest sensitivity for WIMP masses above a few~\gev.  However, in order to probe operators dependent on other nuclear properties, the complementarity of the three G2 target elements merits further investigation.

\begin{figure*}[btp!]
\begin{center}
\includegraphics[width=\textwidth]{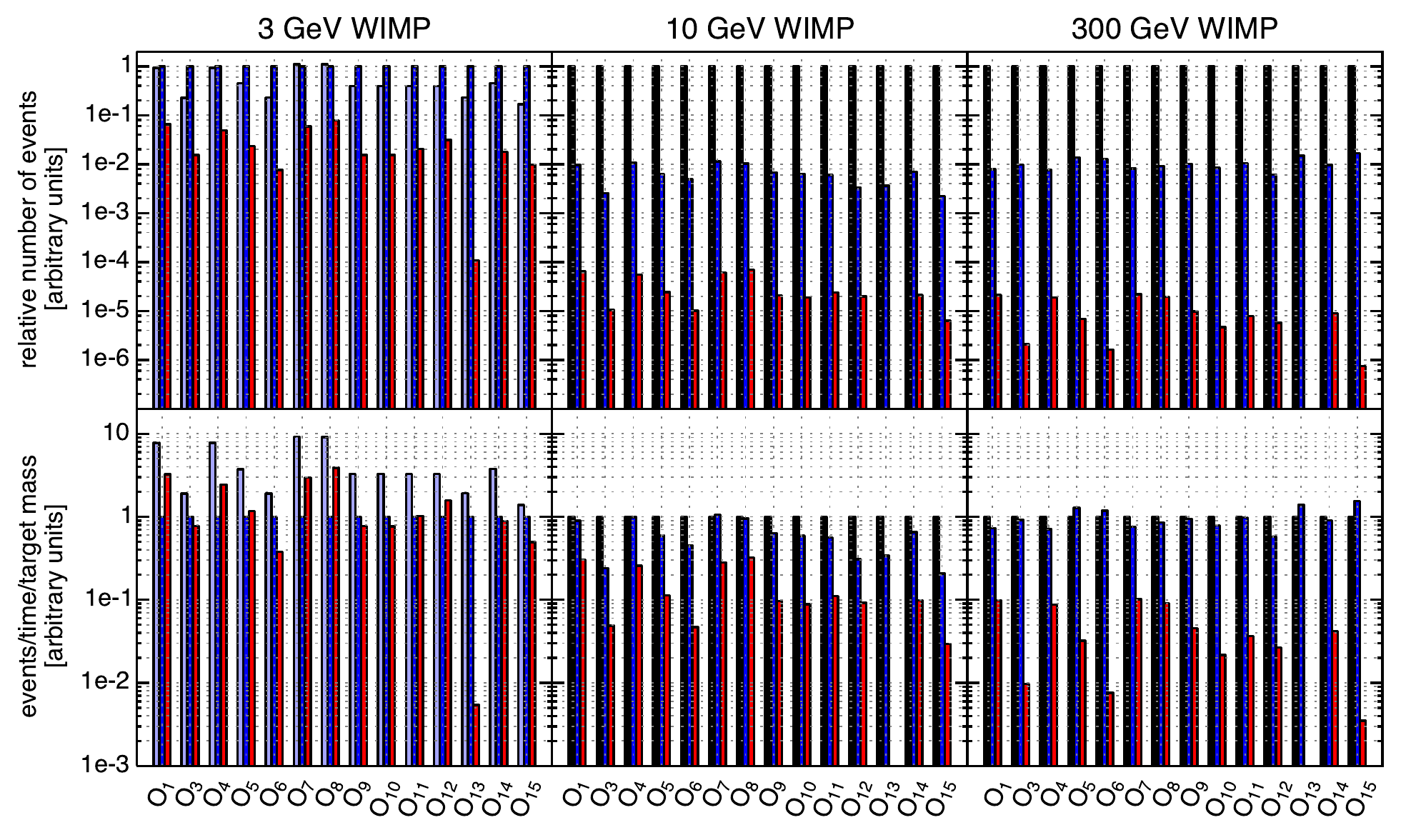}
\caption{Relative event rates for LZ (black), SuperCDMS SNOLAB Ge iZIP (blue), and SuperCDMS SNOLAB Si (red), normalized to 1 observed event in SuperCDMS Ge (3\,\gev) or LZ (10, 300\,\gev).  From left to right are shown the rates for a 3, 10, and 300\,\gev~WIMP, assuming isoscalar interactions and the standard Maxwellian halo model.  The 3\,\gev~case also shows the rates from SuperCDMS SNOLAB Ge high-voltage (light blue), which has similar parameters to SuperCDMS Si high-voltage, but a target mass of 6 kg.  The top row shows cumulative event rates, while the bottom row shows events per time per target mass.  True interaction strengths may differ from this calculation since the interaction may proceed via a linear combination of operators.}
\label{fig:G2_evt_rate}
\end{center}
\end{figure*}

When considering the possible observations the G2 experiments may make, the difference in experimental parameters such as detector mass and trigger threshold must also be taken into account.  The proposed LZ detector will have a 5600\,kg fiducial mass of xenon, while SuperCDMS will be operating 57\,kg of germanium and silicon.  Figure \ref{fig:G2_evt_rate} shows the relative event rates for the three G2 target elements assuming scattering proceeds via a single isoscalar EFT operator.  This figure only shows the relative WIMP rates for the G2 experiments; background rates are not taken into consideration.  Note that the true interaction, which may come from a linear combination of operators, could enhance or suppress these rates.

We normalize the event rate so that SuperCDMS Ge observes one event for a given operator in the 3\,\gev~case and LZ observes one event for a given operator in the 10 and 300\,\gev~cases.  The LZ rate (black) assumes a 5600\,kg fiducial mass, an exposure of 1000\,days, a 100\% trigger efficiency between 1 and 30 keV$_{\mbox{nr}}$, and a flat 50\% nuclear-recoil selection efficiency.  The SuperCDMS Ge rate (blue) assumes 50\,kg of germanium operating in standard iZIP mode \cite{2013iZIPdiscrimination}, an exposure of 1000 days, and a 100\% trigger efficiency between 0.5 and 100\,keV$_{\mbox{nr}}$, and a flat 60\% combined fiducial-volume and nuclear-recoil selection efficiency.  The SuperCDMS Si rate (red) assumes 1\,kg of silicon and an exposure of 1000\,days.  Since the silicon detectors will be operated in high-voltage mode \cite{cdmslitePRL}, the trigger threshold will be much lower, so we assume a 60\% combined trigger and fiducial-volume efficiency up to 50\,keV$_{\mbox{nr}}$, with a trigger threshold of 70\,eV.  We also plot the event rate for SuperCDMS Ge high-voltage (light blue) for the 3\,\gev~WIMP case.  For the SuperCDMS Ge high-voltage detectors, we assume a target mass of 6\,kg, trigger threshold of 80\,eV, and all other parameters identical to SuperCDMS Si high-voltage.

\begin{figure*}[htb!]
\begin{center}
\includegraphics[width=\textwidth]{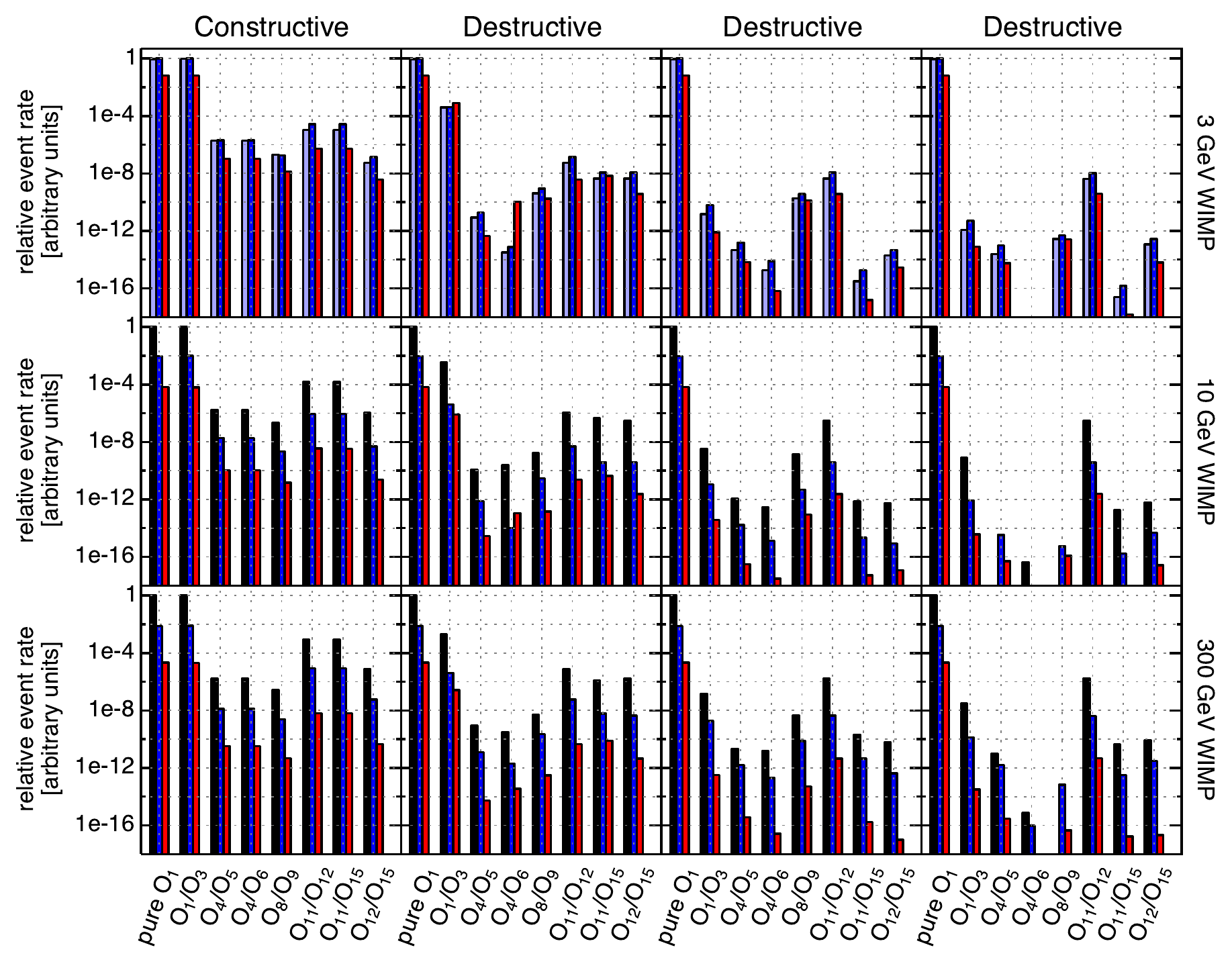}
\caption{Event rate suppression relative to $\mathcal O_1$ scattering in LZ (black), SuperCDMS SNOLAB Ge iZIP (blue), and SuperCDMS SNOLAB Si (red) for interference in germanium, with interference ranging from constructive (left) to maximally destructive (right), as determined by the magnitude of the corresponding eigenvalue.  The rate for SuperCDMS SNOLAB Ge high-voltage (light blue) is shown for the 3\,\gev~case where LZ sees no events above threshold.  The seven operator-operator interference cases are shown, as well as pure isoscalar $\mathcal O_1$, which is used as a reference point.}
\label{fig:G2_interference}
\end{center}
\end{figure*}

Though silicon, germanium, and xenon have similar nuclear properties (e.g.,~all three have isotopes with unpaired neutrons), the variation in the event rate across operators and target elements is large.  For the 3\,\gev~case, the strength of the silicon signal relative to the germanium signal varies by three orders of magnitude, depending on the operator assumed.  The signal in LZ is very close to zero for such a low-mass dark matter particle because the velocity required for a 3\,\gev~WIMP to deposit energy above the assumed 1 keV threshold is greater than the galactic escape velocity.  However, for WIMP masses above a few \gev, LZ's exposure, which is approximately 100 times larger, leads to event rates that are enhanced by approximately the same factor.  In addition, the relative rate for SuperCDMS Si HV becomes smaller at higher masses, since, by design, it is mainly sensitive to the small energy depositions produced by low-mass WIMPs.  

To examine the effects of the different possible interactions for experiments with similar fiducial masses, we also plot the event rate per time per target mass (Fig.~\ref{fig:G2_evt_rate}, bottom).  Here, we see that both Ge and Si SuperCDMS detectors operating in high-voltage mode are more sensitive to low-mass WIMPs because of their lower thresholds.  In particular, the germanium high-voltage rate per kg day (light blue) is nearly an order of magnitude larger than the standard germanium iZIP rate (blue) for certain operators.  For higher masses, the rates for xenon (black) and germanium are comparable within an order of magnitude, but the nuclear properties of silicon (red) make it less sensitive to these interactions.  In addition, SuperCDMS Ge sees a modest enhancement to the overall event rate at high WIMP masses where the distribution of events extends beyond the assumed 30 keV$_{\mbox{nr}}$ upper limit for LZ.  This effect is most prominent for operators such as $\mathcal O_3$ and $\mathcal O_{15}$, which have a $q^2$ dependence that suppresses the rate at low energies, though it is not enough to overcome the effects of LZ's larger target mass in the total number of events.

The variation in signal strength across target elements in this effective field theory solidifies the case for using multiple targets to detect dark matter.  Previous work has shown that complementary target elements can break the degeneracy between the standard spin-dependent and spin-independent operators \cite{2007PhRvL..99o1301B,2013JCAP...07..028C}, and others have shown that this concept can also be applied to the larger EFT parameter space \cite{2014JCAP...09..045C}.  Such considerations are particularly important when incorporating the effect of interference on the event rate.  Because of the presence of both isospin interference and operator-operator interference, there are many combinations of interactions that may greatly suppress the event rate for one particular element.  Even if a single experiment sees no signal due to interference effects, a complementary target with different nuclear properties may still observe events.

To demonstrate the effect of interference on the relative event rate, we determine regions of extremal interference in germanium using the principal component analysis method detailed above.  The event rate suppression relative to $\mathcal O_1$ for the three G2 experiments for germanium constructive interference and destructive interference are shown in Fig.~\ref{fig:G2_interference} for WIMPs with masses of 3, 10, or 300~\gev, assuming the standard Maxwellian halo model and the same experimental parameters as in Fig.~\ref{fig:G2_evt_rate}.  Again, this figure does not consider the relative background rates for the three experiments.  We consider all seven possible cases of four-dimensional operator-operator interference.  The sum of the squares of the EFT coefficients is equal for all cases presented; however, because of the relative strength of various operators and the presence of interference, the rate can be suppressed by many orders of magnitude.  We characterize the interference using the magnitude of the eigenvalue:  the largest eigenvalue corresponds to the maximally-enhanced event rate, while small eigenvalues correspond to varying levels of destructive interference.

The relative event rates in Fig.~\ref{fig:G2_interference} indicate that constructive interference can only modestly enhance the event rate.  In the case of $\mathcal O_1 /\mathcal O_3$ interference, the maximal rate is only $\sim1.5$\% larger than the pure $\mathcal O_1$ rate.  For operators such as $\mathcal O_4$ that depend on the spin of a nucleon in the nucleus, the enhancement relative to the respective isoscalar operator tends to be slightly larger.  In particular, the constructive interference eigenvector for $\mathcal O_4 /\mathcal O_5$ and $\mathcal O_4 /\mathcal O_6$ interference corresponds to WIMP-neutron spin-dependent scattering and is approximately a factor of 2 larger than the isoscalar $\mathcal O_4$ rate.

Since germanium, silicon, and xenon have similar properties, the event rate in SuperCDMS and LZ is suppressed equally for most interference cases.  However, there are a few notable exceptions.  From Fig.~\ref{fig:G2_evt_rate}, we see that for a 3\,\gev~WIMP interacting via a pure isoscalar operator, the event rate in SuperCDMS Si high-voltage tends to be at least an order of magnitude smaller than the rate in SuperCDMS Ge.  When interference is considered, the rate in silicon may become equal to or larger than that in germanium.  As an example, the $\mathcal O_1 /\mathcal O_3$ right-most destructive interference case in Fig.~\ref{fig:G2_interference} corresponds to maximal $\mathcal O_1$ isospin violation in germanium ($f_n/f_p \sim -0.8$) as discussed in \cite{2013PhRvD..88a5021F}.  For this choice of coefficients, the rate in xenon and germanium is suppressed relative to pure isoscalar $\mathcal O_1$ scattering in that target by a factor of $\sim$$500$ and $\sim$$2000$, respectively, while the rate in silicon is suppressed by a factor of $\sim$$100$.  A second instance of this suppression is seen for $\mathcal O_4/\mathcal O_6$ interference at 3\,\gev~in the second plot from the left: the rate in both silicon and germanium is suppressed, but the suppression in germanium is much larger, leading to a greater number of events observed in silicon.

 In addition, there exist several cases for higher WIMP masses where the rate in LZ is smaller than that in SuperCDMS Ge, despite LZ's 100$\times$ larger exposure.  Maximal destructive interference (right-most plot) for $\mathcal O_4/\mathcal O_5$ and $\mathcal O_8/\mathcal O_9$ suppresses the event rate in xenon enough that SuperCDMS will see orders of magnitude more events than LZ, even for larger WIMP masses where LZ typically has an advantage.  For additional interference cases the rate in LZ is less than an order of magnitude larger than that in SuperCDMS Ge.  Although the cases presented here are arguably fine-tuned, the existence of regions of parameter space where interference suppresses the rate in one experiment by orders of magnitude relative to another further supports the need for multiple experiments which use a variety of target elements.

\section{Conclusions}

The interaction between dark matter particles and nuclei might be much more complicated than direct detection experiments have typically assumed. The inclusion of new operators within the framework of an EFT might have profound consequences for current and proposed experiments. As a result, in this richer parameter space, data from multiple experiments with different targets is essential in order to determine the precise nature of the interaction. In addition, when modeling dark matter signals, experiments must consider how an interaction due to an arbitrary EFT operator can affect the energy distribution of dark matter events.

The importance of using multiple target elements to constrain dark matter interactions can already be seen when plotting limits from current experiments.  As we have shown, the differences in target element properties lead to variations in the shape of the interaction strength versus mass limit curve.  In addition, a combination of target elements can produce better constraints on dark matter, especially when considering multiple dark matter interactions and the possibility of interference.  This complementarity of different target elements will become increasingly important in the case of a statistically significant detection.

The additional interactions introduced by the EFT formalism become especially significant when experiments use statistical techniques which rely on assumptions about the shape of the dark matter recoil spectrum to distinguish between background and a potential dark matter signal.  Machine learning techniques, such as the boosted decision tree used in the SuperCDMS Soudan result \cite{R133_LT}, and likelihood analyses, such as the one performed on CDMS~II low-energy data \cite{CDMS-II_likelihood}, require accurate models of both the signal and the expected background.  So far, direct detection experiments have focused primarily on building accurate models of their expected backgrounds, while assuming a simple signal model.  However, mis-modeling the signal could also have significant consequences.  If a WIMP signal that does not conform to the standard spin-independent assumptions is present in the data, it could produce unknown effects on the final result because it may not match either the signal or the background model.  In the case of algorithms such as the optimum interval method that compare the observed events to the expected WIMP spectrum but do not attempt to subtract background, WIMP signal events may be interpreted as background, leading to limits that are too strict.  

These considerations become especially important as the community moves forward with the proposed G2 experiments.  SuperCDMS SNOLAB and LZ will have unprecedented sensitivity to dark matter scattering for a wide range of WIMP masses, and the combination of target elements allows one experiment to verify a potential signal seen by the other.  However, the variation in signal strengths across EFT operators and experimental target elements could lead to experimental results that appear to be in conflict under the standard dark matter assumptions.  In particular, interference between operators can suppress the relative event rates by several orders of magnitude for germanium, silicon, and xenon.  If the true dark matter interaction includes such interference, it is possible that one experiment will observe a statistically significant signal while the other does not.  The effective field theory framework can account for such apparent inconsistencies, and, in the event of a statistically significant signal, it will pave the way for future likelihood analyses to determine the nature of the dark matter interactions.

\section{Acknowledgements}

The authors gratefully acknowledge Liam Fitzpatrick, Wick Haxton, and Tim Tait for helpful conversations.  This work is 
supported in part by the National Science Foundation, by the United States Department of Energy, by 
NSERC Canada, and by MultiDark (Spanish MINECO). Fermilab is operated by the Fermi Research Alliance, 
LLC under Contract No. De- AC02-07CH11359. SLAC is operated under Contract No. DE-AC02-76SF00515 with 
the United States Department of Energy.



\bibliographystyle{apsrev4-1}
\bibliography{biblio_EFT_OI}

\clearpage


\section*{Supplementary material}

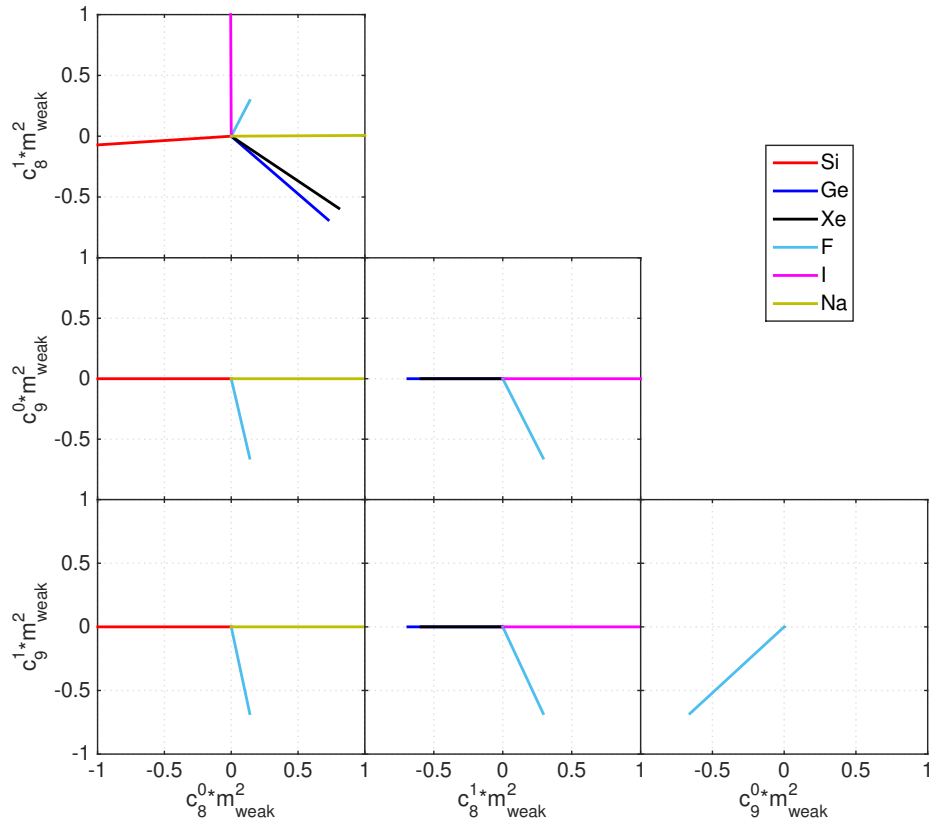

Figure 1: Constructive interference eigenvectors for 4D $\mathcal{O}_8/\mathcal{O}_9$ interference. (Published Fig. 6)

In analyses undertaken after publication of the original article, a minor coding error (bug) has been identified in the utilized code. This bug is in the MATLAB code used to calculate detector target form factors and was identified by disagreement between the MATLAB code and the original Mathematica code [1]. The effects of this bug only occur in the interference between relevant operator pairs, such as operators 8 and 9 ($\mathcal{O}_{8/9}$). Any calculations using non-interfering pairs of operators or single operators are unaffected by the bug. The affected figures published in the original paper are reproduced in their corrected form here. The changes in these figures are relatively minor.

The published Fig. 6 (Fig. 1 here) is one of the three affected plots. The directions of the constructive interference eigenvectors for each target in the 4D interference between operators 8 and 9 change, but the conclusion drawn from the figure does not: Fluorine is still the only target that is sensitive to the isoscalar and isovector components of operator 9.

The published Fig. 7 (Fig. 2 here) is also affected by the coding bug. Because the direction of each constructive interference eigenvector changed, the point in the 4D parameter space at which this figure is made changed as well. For the new directions of the eigenvectors, the shapes of the corresponding event rates are very similar to those originally published with only minor changes. The main message of these plots does not change: For a given target's constructive interference eigenvector, the other targets have lower or suppressed event rates in comparison to the chosen target.

Finally, the published Fig. 9 is also affected. Due to its complexity and the CPU time required to reproduce it, this figure was not regenerated after correction. However, as with the other two figures, the changes in Fig. 9 would be minor and the main message remains unaffected: The event rates in different targets could be different by orders of magnitude depending on the possible constructive or destructive interferences among different operators.

[1] N. Anand, A. L. Fitzpatrick, and W. C. Haxton, Phys. Rev. C **89**, 065501 (2014), arXiv:1308.6288 [hep-ph].

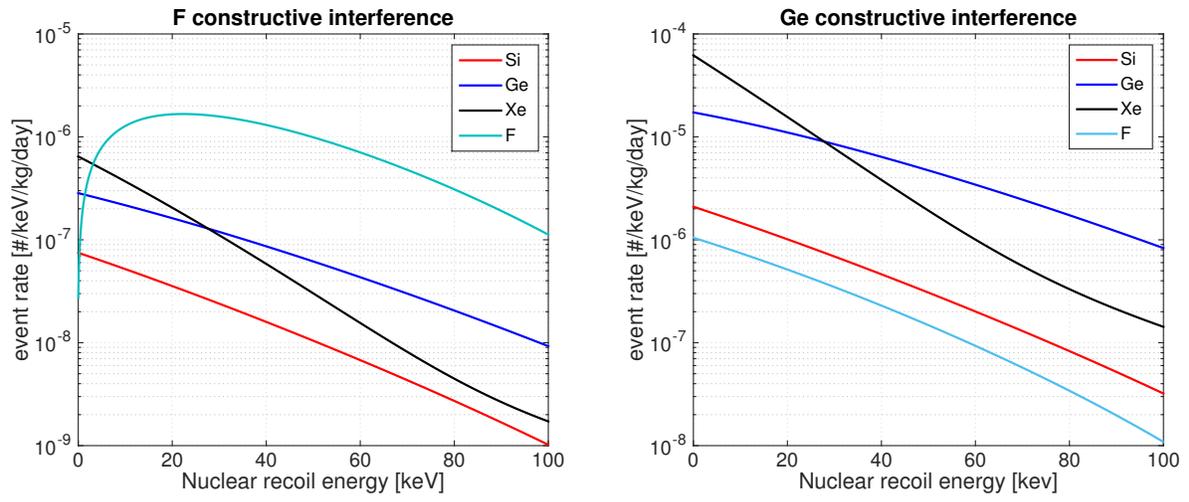

Figure 2: Differential event rate evaluated at the $\mathcal{O}_8/\mathcal{O}_9$ constructive interference vector from Fig. 1 for fluorine (left) and germanium (right). (Published Fig. 7)



\end{document}